\documentclass[sigconf,10pt]{acmart}
\pdfoutput=1

\usepackage[ruled,vlined]{algorithm2e}
\usepackage{amsmath,latexsym,amsfonts,amsthm}
\usepackage{graphicx}
\usepackage{mathtools}
\usepackage{multicol}
\usepackage{lineno,hyperref}
\usepackage{color,colortbl}
\usepackage{float}
\usepackage{booktabs} 
\usepackage{subcaption}
\usepackage{balance}
\usepackage{multirow}
\usepackage{siunitx}
\usepackage{adjustbox}
\usepackage{pifont}
\usepackage{arydshln}
\usepackage{xcolor}
\usepackage[shortlabels]{enumitem}
{
  \theoremstyle{plain}
  \newtheorem{assumption}{Assumption}
}
{
  \theoremstyle{plain}
  \newtheorem*{problem}{Problem}
}

\SetKwInOut{Input}{Input}
\SetKwInOut{Output}{Output}
\SetKwInOut{Data}{Data}

\newcolumntype{L}{>{\centering\arraybackslash}m{1.5cm}}
\newcolumntype{M}{>{\centering\arraybackslash}m{2.68cm}}

\newcommand{\acronym}{FDMonitor}%

\renewcommand\footnotetextcopyrightpermission[1]{} 
\AtBeginDocument{%
  \providecommand\BibTeX{{%
    \normalfont B\kern-0.5em{\scshape i\kern-0.25em b}\kern-0.8em\TeX}}}

\begin{document}

\title{Two Measure is Two Know: Calibration-free Full Duplex Monitoring for Software Radio Platforms}

\author{Jie Wang}
\affiliation{%
 \institution{Washington University in St. Louis}
 \country{}
 }
\email{jie.w@wustl.edu}

\author{Jonathan Gornet}
\affiliation{%
 \institution{Washington University in St. Louis}
  \streetaddress{P.O. Box 1212}
 \country{}
 }
\email{jonathan.gornet@wustl.edu}

\author{Alex Orange}
\affiliation{%
 \institution{University of Utah}
  \streetaddress{1 Th{\o}rv{\"a}ld Circle}
 \country{}
 }
\email{alex.orange@utah.edu}

\author{Leigh Stoller}
\affiliation{%
 \institution{University of Utah}
 \country{}
 }
\email{stoller@flux.utah.edu}

\author{Gary Wong}
\affiliation{%
 \institution{University of Utah}
 \country{}
 }
\email{gtw@flux.utah.edu}

\author{Jacobus Van Der Merwe}
\affiliation{%
 \institution{University of Utah}
 \country{}}
\email{kobus@cs.utah.edu}

\author{Sneha Kumar Kasera}
\affiliation{%
 \institution{University of Utah}
 \country{}
 }
\email{kasera@cs.utah.edu}

\author{Neal Patwari}
\affiliation{%
 \institution{Washington University in St. Louis}
 \country{}
 }
\email{npatwari@wustl.edu}

\begin{abstract}

Future virtualized radio access network (vRAN) infrastructure providers (and today's experimental wireless testbed providers) may be simultaneously uncertain what signals are being transmitted by their base stations and legally responsible for their violations. These providers must monitor the spectrum of transmissions and external signals without access to the radio itself.
In this paper, we propose \acronym{}, a full-duplex monitoring system attached between a transmitter and its antenna to achieve this goal.  Measuring the signal at this point on the RF path is necessary but insufficient since the antenna is a bidirectional device. \acronym{} thus uses a bidirectional coupler, a two-channel receiver, and  a new source separation algorithm to simultaneously estimate the transmitted signal and the signal incident on the antenna. Rather than requiring an offline calibration, we also adaptively estimate the linear model for the system on the fly.  \acronym{} has been running on a real-world open wireless testbed, monitoring 19 SDR platforms controlled (with bare metal access) by outside experimenters over a seven month period, sending alerts whenever a violation is observed.  Our experimental results show that \acronym{} accurately separates signals across a range of signal parameters. Over more than 7 months of observation, it achieves a positive predictive value of 97\%, with a total of 20 false alerts.
\end{abstract}

\maketitle
\pagestyle{plain} 

\section{Introduction}

Demand for spectrum and flexibility has led to technology trends that make radio access networks (RAN) increasingly dynamic and software driven.  Spectrum sharing systems \cite{Hu2018Full}, e.g., the citizens broadband radio service (CBRS) \cite{FCC-CRBS-Order}, require a RAN to dynamically change frequency bands to coexist with multiple different types of wireless systems.  Infrastructure sharing systems, as proposed in emerging virtualized RAN (vRAN) architectures \cite{Intel2017vran, Foukas2017Network, Liang2015Wireless, Richart2016Resource}, provide the flexibility needed to efficiently meet the needs of multiple heterogeneous service providers. As another example, radio dynamic zones (RDZs) are also likely to use dynamic and spectrum-agile approaches to provide services from a shared infrastructure, and thus require spectrum monitoring technologies \cite{NSF-SII-NRDZ-2022}.  These architectures promise to increase the efficiency of spectrum and infrastructure utilization.

\begin{figure}[tb]
    \centering
    \includegraphics[width=1.1\columnwidth]{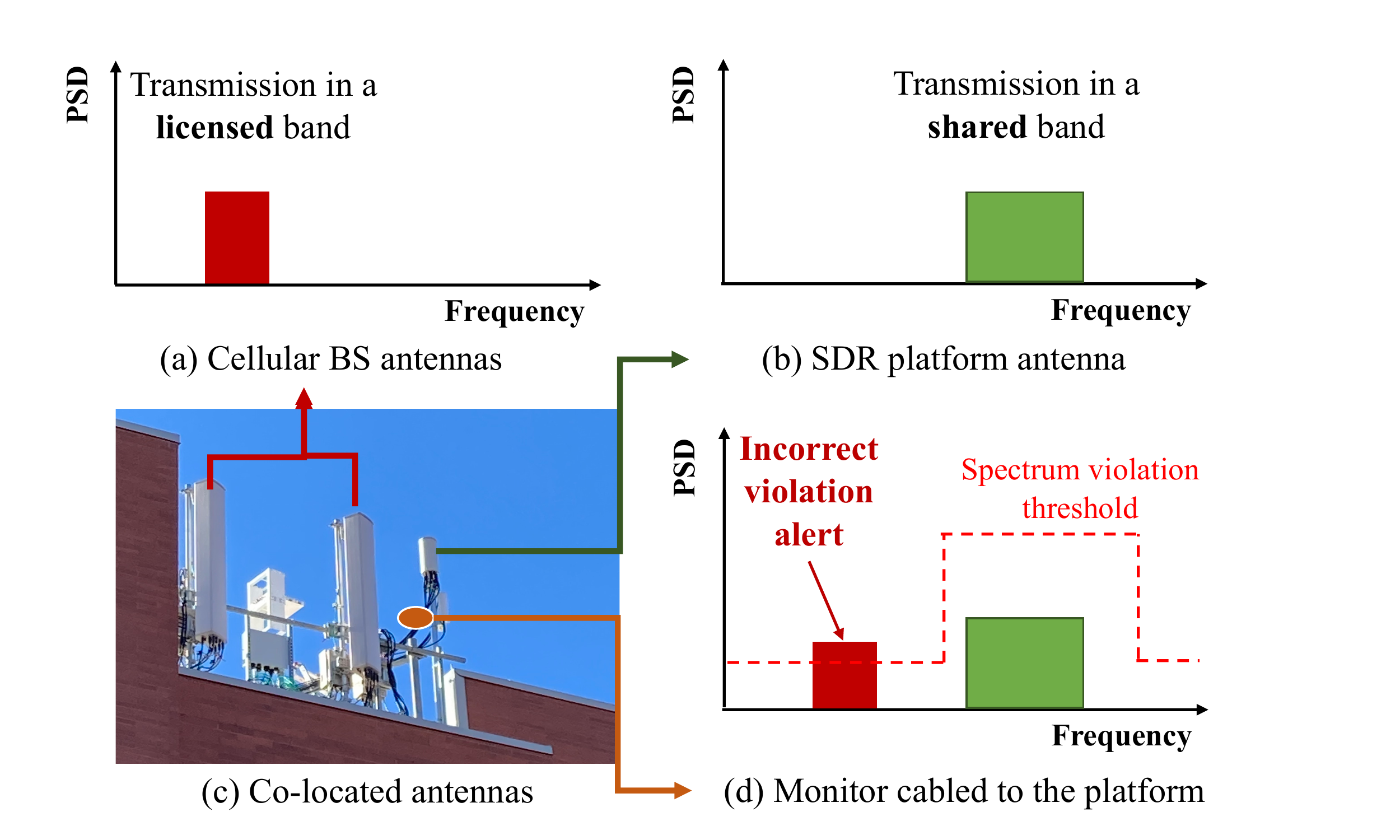}
    \caption{A co-located base station's signal is incident to an SDR platform's antenna.  If the monitor cannot separate the two signals, it may believe that its own SDR's transmitted signal is violating spectrum rules, which would be a false alarm.}
    \label{fig0}
\end{figure}

However, flexibility and dynamism increase the complexity of the enforcement of spectrum regulation.  It is unclear how spectrum sharing systems will identify and handle spectrum violations.  For virtualized RANs, an infrastructure provider (InP) \cite{Liang2015Wireless} can possibly lease a software-defined base station (BS) to a service provider (SP) who might cause a spectrum violation.  In this case, it seems likely that InPs would continue to be legally responsible for spectrum violations as they are today \cite{FCCEnforcementOverview}. Violations may be intentional \cite{Rajendran2020Crowdsourced} or unintentional, caused by erroneous configuration changes, a software bug, or even malware.  Since the transmitted signal spectrum is an interaction between software and analog hardware (power amplifiers, filters, etc.) it is not straightforward to identify problems solely in software. Furthermore, it is advantageous to operate a monitor independent from the BS radio that is leased to a SP.  

While these monitoring problems might seem far in the future, there is a strong need to address these \emph{today} in existing wireless experimental testbeds \cite{PAWR2021, 5GPPP, bristol, ADRENALINE, orbit} that are openly available to researchers.  Users of these testbeds run real-world next-generation wireless experiments with bare-metal access to compute and SDR resources, and thus can transmit with a wide range of frequencies, modulations, and bandwidths.  Users are asked to follow spectrum regulations, but fundamentally, the testbed operator is liable, and must shut down transmitters in violation \cite{FCC-OET}.  In this paper, we design and build \acronym{}, a robust system to reliably conduct this monitoring.

One real-world spectrum violation that we've observed with \acronym{} is when platform users set the gain on the power amplifier too high, in an attempt to maximize the transmit power. Doing so can cause severe harmonics induced by the nonlinearities of the power amplifier. Although the user may be unaware, their high power harmonics can violate out-of-band power limits, and interfere with other licensed users.  Being the source of interference can harm relationships with other (licensed) wireless operators.  We periodically receive inquiries asking if we caused the interference seen by some other licensed operator, and \acronym{} measurements have been critical to demonstrate that our platform was not the cause. 
 
\acronym allows testbed operators to meet two critical requirements:
\begin{itemize}
    \item User monitoring: observe transmissions from its SDR platforms and detect any spectrum violation to ensure compliance. For transmission in shared bands, this requires simultaneously monitoring other users' signals (from external sources).
    \item Spectrum protection: monitor the environmental use of spectrum to observe (to be able to act against) interference sources.
\end{itemize}

We present a full-duplex monitoring system (\acronym{}) \cite{RFmonitor2021github} to achieve simultaneous spectrum monitoring of both the environment and the transmissions by the platform. \acronym{} continuously estimates the signal transmitted by the platform user and external signals. 
Very importantly, \acronym{} addresses a significant challenge that is posed by co-located antennas, as shown in Figure~\ref{fig0}. Consider an SDR platform's antenna deployed on a cellular tower (where platform antennas are deployed to test cellular technologies): although a user is transmitting in an authorized band, an antenna used by a cellular provider $CP$ might transmit at 50~W on the same tower\footnote{Towers and rooftop antenna locations are typically owned by an InP who leases tower space to multiple wireless operators.}. Since an antenna is a two-way device, some of $CP$'s signal, although many dB less than transmitted, impinges on the platform's antenna. If we directly cable a monitor to the platform, it records both transmitted and incident signals, and cannot distinguish them. \textit{In this case the monitor wrongly concludes that the user is transmitting in the band owned by CP and, for spectrum compliance, aborts the user's operation}.

\begin{figure}
    \centering
    \includegraphics[width=1.0\columnwidth]{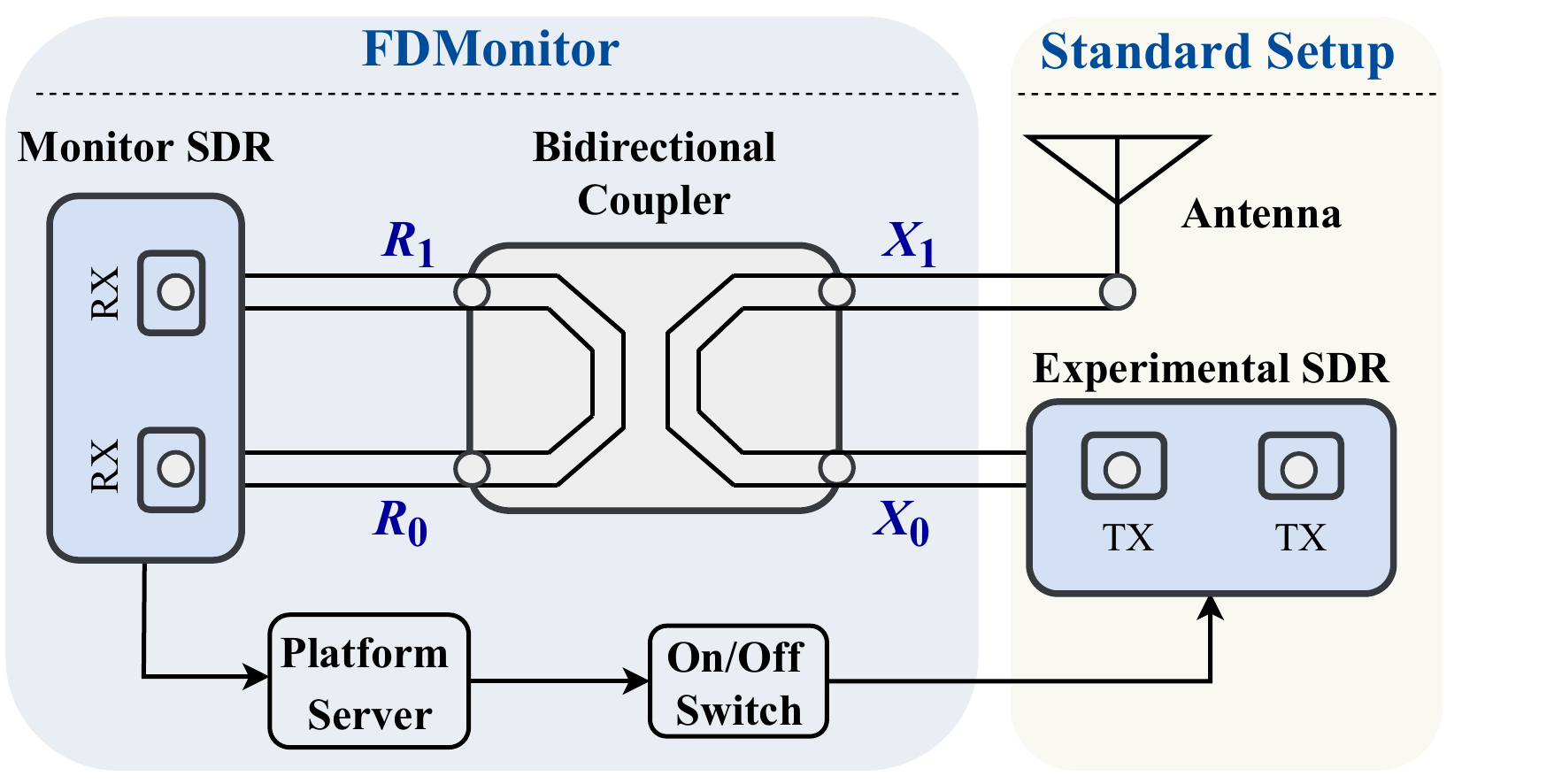}
    \caption{Architecture of \acronym, the proposed full-duplex monitoring system}
    \label{Architecture}
\end{figure}

\textbf{Hardware design of \acronym{}}. The design of \acronym{} addresses the problem that antennas both emit (most of) the platform's transmit signal and receive signals from external co-channel sources. As shown in Figure~\ref{Architecture}, we use a bidirectional coupler to measure the forward and backward traveling signals on the RF path, in two \textit{different} linear combinations, 
which we call $R_0$ and $R_1$. Here, $R_0$ receives more of the signal from the platform transmitter than $R_1$, whereas, $R_1$ receives more of the incident signal than $R_0$. 

Unfortunately, $R_0$ and $R_1$ are misleading estimates of the actual transmitted signal (denoted by $X_0$) and the incident signal (denoted by $X_1$), respectively. The challenge is that a wideband bidirectional coupler does not perfectly isolate these two signals --- the overall system can provide only 10--15 dB difference in the power of one source between the two coupled outputs. This is because RF subsystems are not perfectly matched over the wide bandwidths across which frequency-agile SDR platforms must be able to operate. 
Counterintuitively, the platform's transmitted signal and the incident signal are carried in \textit{both} directions on the RF chain, so a \textit{directional} coupler can only do so much. 

\textbf{Source Separation in \acronym{}}. Separating the two signals (transmitted and incident), without knowing anything about either signal, becomes our significant problem.
An insight we use is that the transmitted signal should be independent of the signal incident to the antenna, and further, most digital communication signals are non-Gaussian.  This insight allows us to apply blind source separation algorithms like independent component analysis (ICA) to separate the transmit and incident signals. 
However, ICA results in scaling and permutation ambiguities. Our source separation algorithm uses the mixing matrix inverse to recover the scale, which gives us accurate signal power levels. Correlation coefficients \cite{AKOGLU2018correlation} and the maximum power of the coupler's two receiver chains are used in the algorithm to correctly identify the two estimates as ``transmitted'' or ``incident''.

We implement \acronym{} and deploy it on 19 SDRs available to researchers on an open wireless research testbed, and thoroughly evaluate its performance with different RF settings (carrier frequency, bandwidth, transmit power and signal type) to verify that \acronym{} operates correctly.  
\acronym{} has been running continuously for over 7 months on the testbed, alerting administrators of any spectrum violation, who may then turn off the user's transmitter. We investigate each violation and find that \acronym{} has achieved a $97\%$ positive predictive value over this 7 month period.

\section{Related Work} \label{sec:related}
Full-duplex monitoring of the shared SDR platforms is at the intersection of relevant research on (i) vRAN, (ii) full-duplex communication, and (iii) spectrum sensing.


\subsection{Virtual RAN} 
Next-generation virtual RAN architectures are rapidly evolving.  We describe vRAN here to address its novel needs for spectrum monitoring.

\textbf{vRAN Architecture}.
In a conventional distributed RAN (D-RAN), every BS is composed of a baseband unit (BBU) and remote radio heads (RRHs), all located at the same cell site \cite{Habibi2019RAN}. RRHs interface with the antenna on one end, and the BBU via fiber on the other. RRHs have radio frequency (RF) capabilities like power amplification and sampling, whereas BBUs host baseband functions for further processing.

In comparison to D-RAN, vRAN's main idea is network function virtualization (NFV) in which network functions are decoupled from the physical hardware \cite{Mijumbi2016NFV}. BBUs of all BSs, based on NFV principles, are not only centralized, but also virtualized as a cloud. They are deployed on the same commodity server hardware and can be created dynamically when changes are needed over frequency, time and space \cite{Intel2017vran}. RRHs, instead, remain at the cell towers yet support flexible functions via SDR hardware. 

\textbf{The Business Model of vRAN}.
Multiple roles are needed in vRAN. 
Different business models map these roles with actual entities \cite{Costa2013RadioAN, Liang2015Wireless, Jiang2021Decentralized, Kitindi2017WirelessNV}. 
Roles and functions include:
\begin{itemize}
    \item \textit{Infrastructure providers (InPs)} deploy and own the physical infrastructure including RAN, fronthaul/backhaul transport, core networks and etc. BSs are provided to SPs as a service by splitting them into slices. InPs are required to detect and disable spectrum violations.
    \item \textit{Virtual network providers (VNPs)} can operate on licensed or unlicensed spectrum and create virtual network resources for SPs to use. VNPs may or may not own spectrum resources.
    \item \textit{Service providers (SPs)} lease BSs from InP(s) and virtualized network resources from VNPs. In doing so, they can provide services to their subscribers.
\end{itemize}

\textbf{Experimental Wireless Testbeds}.
Experimental wireless testbeds can be seen as an abstraction of a vRAN today. The testbed providers are the InPs responsible for managing the deployed infrastructure and detecting spectrum violations. The VNPs, include various entities, such as  spectrum access system (SAS) operators for the CBRS band \cite{sas2020cbrs} or mobile network operators for cellular bands \cite{Liang2015Wireless}. Researchers are the SPs requesting shareable physical and network slices for testing new wireless technologies.

Recent wireless infrastructures include: (i) POWDER, COSMOS and AERPAW testbeds \cite{Breenwintech20, Raychaudhuri2020COSMOS, Marojevic2020AERPAW} as part of the U.S.\ NSF's platforms for advanced wireless research (PAWR) program \cite{PAWR2021}; (ii) 5G-VINNI, 5G-EVE and 5GENESIS \cite{5GPPP2021report} as 5G testbeds being developed by joint efforts across Europe; (iii) Bristol Is Open in the UK \cite{Harris2016BIO} and ADRENALINE \cite{Raul2017ADRENALINE} in Spain for exploring 5G, software defined networking and large-scale Internet of things (IoT) techniques.

\subsection{Full-duplex Communication} 
Full-duplex communication \cite{Sabharwal2014InBand, Liu2015InBand, Zhang2016Full, Kolodziej2019InBand} enables simultaneous transmission and reception in the same channel. It requires neither time division duplexing (TDD) or frequency division duplexing (FDD) and thus significantly increases spectral efficiency and network capacity \cite{AKYILDIZ20165G}. Both full-duplex monitoring and full-duplex communication allow co-channel signal differentiation, but the former is focused on monitoring spectrum use, while the latter is for bidirectional communication.
Self-interference (SI) \cite{Amjad2017Full}, i.e., the contamination of the received signal with the transmitted signal, is the biggest challenge for both. 
Proposed SI cancellation methods can be classified as: (i) propagation-domain, (ii) analog-domain, and (iii) digital-domain. 

\textbf{Propagation-domain SI cancellation} methods isolate the transmit and receive chain carefully for electromagnetically suppressing the SI before it shows up in the analog circuitry \cite{Sabharwal2014InBand}, via path loss enhancement \cite{Kolodziej2019InBand}, cross-polarization \cite{Aryafar2012MIDU}, transmit beamforming \cite{Riihonen2011Mitigation}, or a circulator \cite{Bharadia2013Full}. 

\textbf{Analog-domain SI cancellation} methods
subtract a copy of the transmitted signal from received signals in the analog receive chain \cite{Sabharwal2014InBand}. The methods can be classified as non-adaptive \cite{Duarte2012Experiment} and adaptive \cite{Jain2011Practical} depending on whether time-varying environment is taken into consideration. 

\textbf{Digital-domain SI cancellation} 
methods cancel SI from quantified received signals after the analog-to-digital converter (ADC) \cite{Liu2015InBand}. A digital domain SI canceller first builds a baseband-equivalent model using the known transmit signal to capture everything between the DAC and ADC \cite{Kolodziej2019InBand}. It then estimates linear and nonlinear components of SI based on the modeled channel to cancel the known transmit signal. 


Compared to these SI cancellation approaches, \acronym{} does not know the transmitted signal, and thus we cannot simply subtract it to estimate the other signal. \acronym{} first works in the propagation domain with its bidirectional coupler to enhance isolation between transmitted and incident signals. However, the isolation is insufficient due to matching across a very wide band of SDR operation. \acronym{} applies a blind frequency-domain source separation algorithm in the digital domain for further signal separation.

\subsection{Spectrum sensing} 
A shared platform's transmission could be monitored by repurposed spectrum sensing. Spectrum sensing \cite{ZengYonghong2010ARoS} was proposed to sense primary users for opportunistic spectrum reuse, but it is, in essence, an approach to remotely detect transmission. Repurposed spectrum sensing can be categorized as (i) direct sensing, and (ii) cooperative sensing.

\textbf{Direct sensing}  uses one node to locally sense a user's transmission \cite{Yucekasurvey2009}. It can be 1) transmitted signal prior-based sensing and 2) blind detection. The first type includes likelihood ratio test \cite{ZengYonghong2010ARoS}, cyclostationarity detection \cite{Han2006Spectral}, waveform based sensing \cite{Tang2005Some}, and matched filtering \cite{Cabric2006Spectrum}. These methods use priors including signal distributions, cyclostationarity, preamble and pilot pattern of the transmitted signals to be correlated with the received signal for signal presence detection. Blind detection, in contrast, does not require a prior \cite{Axell2012Spectrum}. It includes energy detection (ED) \cite{Atapattu2011Energy} or eigenvalue/covariance based detection \cite{Awin2019Blind}. ED measures the direct energy output whereas the latter uses the covariance matrix as an indicator of the received signal strength for presence classification.

\textbf{Cooperative sensing}  utilizes measured signals sharing among collaborative radios to enhance the transmission sensing performance \cite{ZengYonghong2010ARoS}. The spatial distribution of multiple nodes effectively avoids hidden node problems and ameliorates degradation due to multipath fading and shadowing \cite{Ghasemi2008Spectrum}. While cooperative sensing can be centralized, distributed and cluster-based \cite{AKYILDIZ2011Cooperative}, the fundamental sensing method is still direct sensing.

The above  sensing techniques can be repurposed to monitor targeted transmissions of the shared SDR platform. However, it is unrealistic for us to require priors on the transmitted or incident signals; and blind methods will not be able to separate co-channel signals. 
In comparison, \acronym{} can precisely separate and identify both the transmitted signal and incident signal, even if they are on the same channel.  It does use two measurements, thus like cooperative methods, it benefits from redundant measurements.

\textbf{Bidirectional sensing}. The preliminary work described in \cite{Terry2020spectrum} reports on spectrum monitoring using a bidirectional coupler. That method assumes a known system model, and estimates the transmitted signal using the model inverse.  However, system model calibration requires time-intensive manual effort. Furthermore, weather changes result in system variations which, if not recalibrated, reduce the separation performance. Experimentally, we also find the approach cannot sufficiently separate the transmitted and incident signals when they overlap in the frequency domain.  In comparison to \cite{Terry2020spectrum}, \acronym{} provides several new benefits: 1) it is robust across signal type, carrier frequency, bandwidth and transmit power, 2) it enables mixing matrix estimation on the fly without system calibration, and 3) in addition to estimating the transmit signal, it \textit{also} estimates the incident signal, which enables full-duplex monitoring.

\section{System Design}\label{sec:model}
\begin{figure}
    \centering
    \includegraphics[width=1.0\columnwidth]{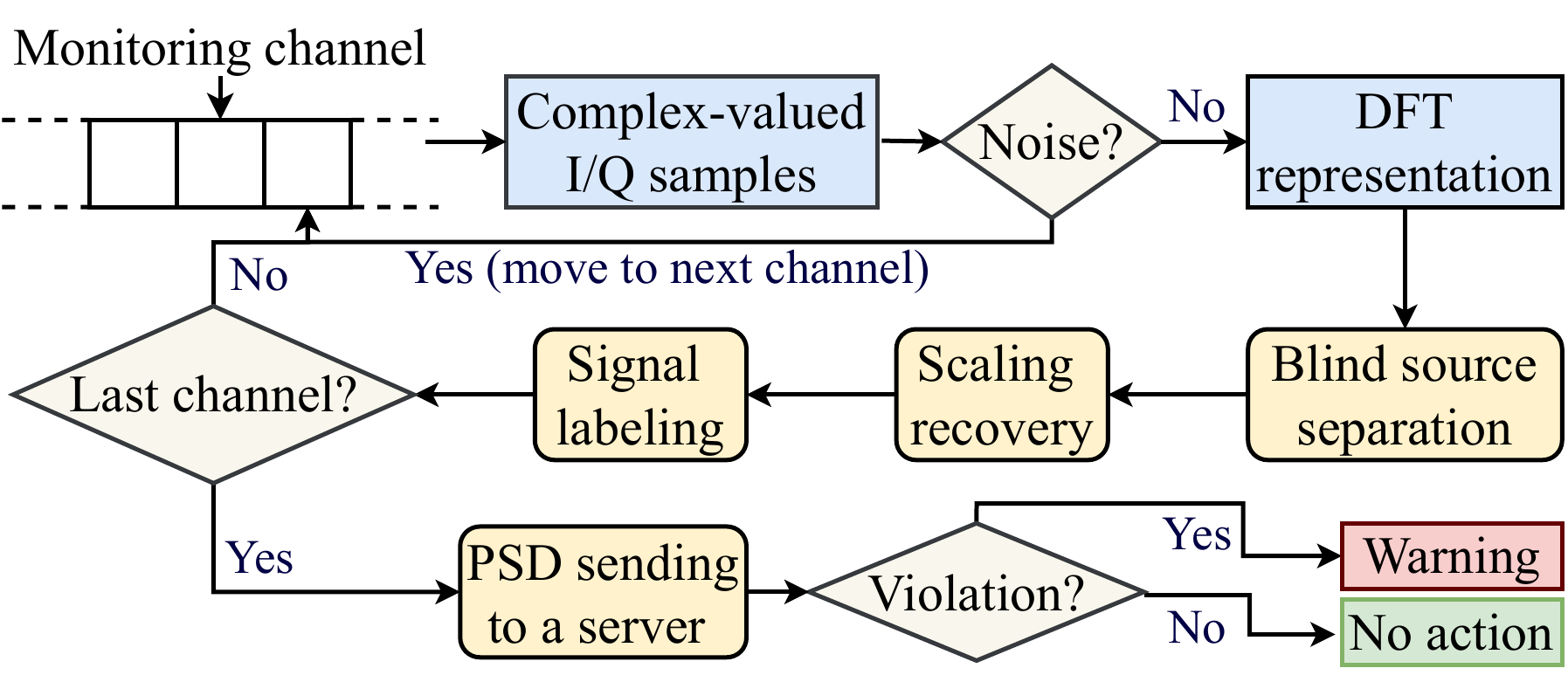}
    \caption{System components of \acronym{}}
    \label{diagram}
\end{figure}

We describe the design of \acronym{}, as shown in Figure~\ref{diagram}, that can separate signals without a signal prior. The inputs to \acronym{} are the in-phase and quadrature (IQ) sampled signals at the two monitoring receiver channels, labeled $R_0$ and $R_1$ in Figure~\ref{Architecture}. Whenever the power in either channel is higher than the noise floor, we use the proposed algorithm to separate the signal into two sources. Given the power spectral density (PSD) limits defined by the platform operator, \acronym{} determines whether a violation occurred and reacts accordingly. In the following sections, describe the components of \acronym{} shown in Figure \ref{diagram}.

\subsection{Problem Formulation}

The overall goal of \acronym{} is to monitor the entire range of the platform, in our case 100--6000~MHz. The monitor samples from one band at a time, each with an RF bandwidth limited by the capability of the monitoring device (in our case, 27.65~MHz).  We describe, without loss of generality, how \acronym{} operates on a single frequency band. 

\acronym{} collects bidimensional samples $r_i(n)$ for sample $n=0, 1,\ldots N-1$ from ports $i=0,1$. Upon referring the source signals as $x_i(n)$ with $i=0,1$, we describe the bidimensional observations of the form:
\begin{equation}\label{eq:timemixture}
    \begin{aligned}
    &\boldsymbol{r}(n) = \boldsymbol{A} \boldsymbol{x}(n) + \boldsymbol{v}(n), \quad \boldsymbol{v}(n) \sim \mathcal{CN}(\boldsymbol{0}, \sigma^2\boldsymbol{I})\\
    &\boldsymbol{r}(n) = \begin{bmatrix}
      r_0(0) & \ldots & r_0(N-1) \\
      r_1(0) & \ldots & r_1(N-1) 
    \end{bmatrix} \in \mathbb{C}^{2\times N}\\
    &\boldsymbol{x}(n) = \begin{bmatrix}
      x_0(0) & \ldots & x_0(N-1) \\
      x_1(0) & \ldots & x_1(N-1) 
    \end{bmatrix} \in \mathbb{C}^{2\times N}.\\
    \end{aligned}
\end{equation}
Here $\boldsymbol{v}(n)$ is the zero-mean, uncorrelated additive Gaussian noise. We now present the assumptions based on (\ref{eq:timemixture}) and discuss the major problem that needs to be addressed.


\begin{assumption}\label{as:linear}
    The observations\, $\boldsymbol{r}(n)$ are a noisy instantaneous linear mixture of source signals from $\boldsymbol{x}(n)$, and the system matrix $\boldsymbol{A}\in \mathbb{C}^{2\times 2}$ is assumed to be unknown.
\end{assumption}
Signal mixtures can be instantaneous or convolutive \cite{choi2005blind, pedersen2008convolutive}. We assume the former due to the fact that \acronym{} collects $\mathrm{I/Q}$ samples at the same time. We assume no prior knowledge of the system matrix $\boldsymbol{A}$ so that calibration measurements are not required; and because weather and other changing conditions alter $\boldsymbol{A}$ in practice. Instead, \acronym{} can adaptively estimate the linear model on the fly. 

\begin{assumption}\label{as:sources}
    Source signals $x_0(n)$ and $x_1(n)$, for $n=0, 1, \ldots N-1$, are mutually independent and unknown. At most one of the source signals is Gaussian distributed.
\end{assumption}
As the transmitted and incident signals, are from different sources: the SDR platform and outside world. One signal does not affect the other, leading to mutual independence. \acronym{} does not know what source signals are, as their properties are designed by platform users, and may even be proprietary and confidential. 

Additionally, the noise power threshold in \acronym{} ensures that we do not operate source separation when both signals are purely noise.  Given that digital signals are mostly non-Gaussian, assumption \ref{as:sources} holds.



Thus we design \acronym{} to solve the following problem:

\begin{problem}\label{prob:formu}
Given assumptions~\ref{as:linear}, \ref{as:sources} and the linear model in (\ref{eq:timemixture}), the challenge is to estimate transmitted and incident signals from the received bidimensional measurements only, which is commonly referred as blind source separation (BSS) for instantaneous linear mixture.
\end{problem}



\subsection{Frequency-domain ICA Modeling}\label{model:FreqICA}
We introduce a well-known BSS technique, independent component analysis (ICA), to address the problem.
ICA approaches, according to \cite{naik2011overview}, are identical to BSS solutions for instantaneous linear mixtures. They assume the same linear mixture framework as in (\ref{eq:timemixture}) and require assumptions~\ref{as:linear} and \ref{as:sources}. 
ICA may be applied in either the time or frequency domain due to the linearity of Fourier transform. 
We adopt the frequency-domain ICA given the spectrum monitoring application of this work. In detail, complex-valued samples $r_i(n)$ for $i=0,1$ are first converted to frequency-domain components via Discrete Fourier Transform (DFT):
\begin{equation}\label{prob:dft}
    R_i(k) = \sum_{n=0}^{N-1} r_i(n) e^{-j\frac{2\pi}{N}kn}, \quad k=0, 1,\ldots, N-1.
\end{equation}

The linear model in (\ref{eq:timemixture}) can then be rewritten in the frequency domain as:
\begin{equation}\label{eq:freqmixture}
    \begin{aligned}
    &\boldsymbol{R} = \boldsymbol{A} \boldsymbol{X} + \boldsymbol{V}, \quad \boldsymbol{V} \sim \mathcal{CN}(\boldsymbol{0}, N\sigma^2\boldsymbol{I}),
    \end{aligned}
\end{equation}
where $\boldsymbol{R} = \begin{bmatrix}
  R_0 & R_1
\end{bmatrix}^T \in \mathbb{C}^{2\times N}$ are the DFT components of raw samples and $\boldsymbol{V}$ in the frequency domain is still Gaussian distributed.  The objective of frequency-domain ICA is to estimate the system matrix $\boldsymbol{A}$ and source signals $\boldsymbol{X}$ in (\ref{eq:freqmixture}) in each monitoring channel.


\subsection{ICA for Source Separation}\label{model:JADE}

ICA methods include two steps for source separation: 1) data pre-processing and 2) contrast  (cost) function optimization for estimating mixing matrix. Data pre-processing consists of data centering and whitening to produce zero-mean, uncorrelated components with unit variance \cite{naik2011overview}. For the second step, different contrast functions that quantify statistical independence have been proposed in the last two decades including cumulants, entropy, and mutual information \cite{Aapo2020independent}. With a selected contrast function, an ICA method can use Jacobi optimization or gradient-based schemes to iteratively estimate the system matrix \cite{cardoso1999high}.

The following ICA algorithms, as applicable to complex-valued signals, are considered in this paper: 1) Fast Independent Component Analysis (FastICA) \cite{Ella2020fast}; 2) Joint Approximate Diagonalization of Eigen-matrices (JADE) \cite{Cardoso1993jade}; and 3) Adaptable Complex Maximization of Nongaussianity (A-CMN) \cite{Novey2006acmn}. All three algorithms are widely used for various applications. However, FastICA and A-CMN require some prior knowledge of the source distribution in order to optimally choose the contrast function \cite{Novey2006acmn}. Although this increases algorithm robustness, this knowledge is not generally available to SDR platform operators who want to accommodate as many wireless techniques as possible. Hence \acronym{} cannot limit the PHY to particular source distributions. JADE, which demands no such prior, is more adaptable to different source conditions and therefore has been used in \acronym{}. The JADE algorithm is shown in Algorithm~\ref{alg:jade}. Note that the separated estimates after ICA have neither the same magnitude as that of raw samples nor correct labeling of ``transmitted'' vs.\ ``incident'', due to two common ICA ambiguities we discuss in the next section.

\SetKwComment{Comment}{/* }{ */}
\begin{algorithm}[ht]
\caption{The JADE algorithm}
\SetAlgoLined
\KwData{DFT components of raw samples $\boldsymbol{R} \in \mathbb{C}^{2\times N}$}
\KwResult{Source estimates $\boldsymbol{\Tilde{X}}$ and the mixing matrix estimate $\boldsymbol{\Tilde{A}}$}
\Comment{Data preprocessing}
    $\boldsymbol{\Bar{R}}(k) = \boldsymbol{R}(k) - \textrm{avg}(\boldsymbol{R}(k))$\Comment*[r]{centering}
    $\boldsymbol{E}, \boldsymbol{D} \gets EVD( \textrm{Cov}\{\boldsymbol{\Bar{R}}(k)\} )$\; 
    $\boldsymbol{\Tilde{D}}=\boldsymbol{D}^{-\frac{1}{2}} \boldsymbol{E}^H,\, \boldsymbol{\Tilde{R}}(k) = \boldsymbol{\Tilde{D}} \boldsymbol{\Bar{R}}(k)$\Comment*[r]{whitening}
\Comment{Contrast function optimization}
    \For { i,j,l,m = 0,1} {
    $Q(\boldsymbol{\Tilde{R}}) = \mbox{Cum}\left(\tilde{R}_i, \tilde{R}_j^H, \tilde{R}_l, \tilde{R}_m^H \right)$ \Comment*[r]{kurtosis}
} 
Compute $\Phi$ = $\left\{\lambda_i, M_i \middle| \; i=0,1\right\}$ as the two largest \{eigenvalue, eigenmatrix\} set of $Q(\boldsymbol{\Tilde{R}})$\; %

Compute a unitary matrix $\boldsymbol{\Tilde{U}}$ by jointly diagonalizing $\Phi$\;

Compute $\Tilde{\boldsymbol{X}}(k) = \boldsymbol{\Tilde{U}}^H \boldsymbol{\Tilde{R}}(k)$\;
Compute $\boldsymbol{\Tilde{A}}=\boldsymbol{\Tilde{D}}^{\#} \boldsymbol{\Tilde{U}}$\Comment*[r]{$\boldsymbol{\Tilde{D}}^{\#}:$ pseudoinverse}

\label{alg:jade}
\end{algorithm}

\subsection{Scaling and Permutation Alignment} \label{S:ScalingAndPermutation}
ICA methods have two common problems: scaling ambiguities and permutation ambiguities. First, ICA solutions are scaled by an unknown constant. Second, the two signal estimates are arbitrarily assigned, thus it is not known which signal was ``transmitted'' which was ``incident''. These ambiguities impose great challenges on violation detection as \acronym{} does not know which estimate to look at for violation detection, and which for environmental spectrum monitoring. 

Assume that the two ICA estimates are each scaled by a multiplicative factor and may be permuted (i.e., swapped). These changes are modeled via the mixing matrix as:
\begin{equation}\label{ambiguity}
    \boldsymbol{\Tilde{A}} \longleftarrow \boldsymbol{\Lambda} \boldsymbol{\hat{A}} \boldsymbol{W},
\end{equation}
where $\boldsymbol{\Lambda}$ is a diagonal scaling matrix. $\boldsymbol{\hat{A}}$ is the ultimate mixing matrix to be obtained, and $\boldsymbol{W}$ is \textit{either} the $2 \times 2$ identity matrix, or if it is permuted, the $2 \times 2$ exchange matrix \(\boldsymbol{J}_2 \overset{\Delta}{=}
\begin{bmatrix}
    0 & 1 \\
    1 & 0 
\end{bmatrix}\).

In the next sections, we first recover the scale via the estimated mixing matrix and address permutation ambiguity using correlation coefficients and power differences.

\subsubsection{\textbf{Scaling Alignment}}\label{S:Scaling}
We observe, from (\ref{ambiguity}), that $\boldsymbol{\Tilde{A}}$ having a norm larger than 1 essentially causes the scaling ambiguity challenge. To recover the scale, we first diagonalize the mixing matrix $\boldsymbol{\Tilde{A}}$ to obtain a complex-valued diagonal matrix $\Delta(\boldsymbol{\Tilde{A}})$:
\begin{equation}\label{equ10}
    \Delta(\boldsymbol{\Tilde{A}}) = 
    \textrm{Diag}\{\boldsymbol{\Tilde{A}}\}.
\end{equation}

The correct power level of the two estimates is then:
\begin{equation}\label{powrecovery}
    \hat{\boldsymbol{X}}(k) = \Delta(\boldsymbol{\Tilde{A}}) \Tilde{\boldsymbol{X}}(k), \, k= 0,1, \ldots, N-1.
\end{equation}

\subsubsection{\textbf{Permutation Alignment}}\label{S:Permutation}
To simply the problem, we first hypothesize the following:
``transmitted'' $\xleftrightarrow[]{} \hat{X}_0(k)$, ``incident'' $\xleftrightarrow[]{} \hat{X}_1(k)$. 
The permutation is incorrect if the hypothesis is false. As we know that $R_0(k)$ from port 0 detects more of the transmitted signal than port 1 does given by \acronym{}'s directionality, we first use the Pearson correlation coefficient \cite{AKOGLU2018correlation} to test the hypothesis:
\begin{equation}\label{equ11}
    \textrm{corr}(|\hat{X}_i|, |\Tilde{R}_j|) = \frac{\textrm{Cov}\{|\hat{X}_i|, |\Tilde{R}_j|\}}{\sigma_{|\hat{X}_i|} \sigma_{|\Tilde{R}_j|}}, \quad i, j = 0, 1,
\end{equation}
where $|\cdot|$ denotes the magnitude of the complex value. $\textrm{Cov}\{\cdot\}$ is the covariance and $\sigma$ is the standard deviation operator, calculated over all frequency samples $k$. We then are able to align the results based on the maximum of the four correlation values:
\begin{equation}\label{equ12}
    \hat{i}, \hat{j} = \operatorname*{arg\,max}_{i,j\in \{0,1\}} \textrm{corr}(|\hat{X}_i|, |\Tilde{R}_j|).
\end{equation}
If indices $\hat{i} = \hat{j}$, regardless of the value, the hypothesis is correct. Otherwise, we multiply $\hat{\boldsymbol{X}}$ by the exchange matrix $\boldsymbol{J}_2$ to swap them back.

The above solution applies to all but source signals of similar power spectra shapes as the correlation coefficients, in this case, are closely high and hence unreliable. As a result, we propose to further align permutation via power maximum between $\Tilde{R}$ and $\hat{X}$ if the correlation coefficients fall out of the 95\% confidence interval. Specifically, If $\Tilde{R}_0$ has more power than $\Tilde{R}_1$, $\hat{X}_0$ should correspondingly have higher magnitude than $\hat{X}_1$. Therefore the $\Tilde{R}_j$ and $\hat{X}_i$ with higher power are matched.

\subsubsection{\textbf{Mixing Matrix Adjustment}}
Certain adjustments of the estimated mixing matrix $\boldsymbol{\Tilde{A}}$ are needed to properly account for the changes we made for scaling and permutation alignment in Section~\ref{S:Scaling} and \ref{S:Permutation}. The final estimated mixing matrix is:
\begin{equation}\label{equ13}
    \hat{\boldsymbol{A}} = \Large(\Delta(\boldsymbol{\Tilde{A}})\Large)^{-1} \boldsymbol{\Tilde{A}} \boldsymbol{W}
\end{equation}
where $\Large(\Delta(\boldsymbol{\Tilde{A}})\Large)^{-1}$ is the scaling recovery matrix and $\boldsymbol{W}$ is the permutation recovery matrix: if permutation occurs, \(\boldsymbol{W}=\boldsymbol{J}_2\), otherwise it is the 2x2 identity matrix. 

\subsection{Performance evaluation}

We propose a new evaluation metric here to quantitatively evaluate the ability of \acronym{} to separate the sources. In particular, we do not want the algorithm to ``blame'' the user for the incident signal, nor do we want to corrupt the incident signal estimate with the user's transmitted signal. The principle is that the power spectral density (PSD) of $\hat{X}_0$ should be high, and much lower in $\hat{X}_1$, over the frequencies containing the transmitted signal. Similarly the PSD of $\hat{X}_1$ should be high, and much lower in the $\hat{X}_0$, over the frequencies where there is an incident signal.

To define the metric, we define a frequency band (set) $\mathcal{B}_{Tx}$ to contain frequency indices in which the node is transmitting; and set $\mathcal{B}_{In}$ to contain frequency indices in which there is incident signal. We compute the average PSDs of these signals in estimated signals $\hat{X}_i(k)$, for $i=0,1$, as:
\begin{eqnarray} \label{E:averagePowerInEstimates}
  P_i[Tx] &=& \frac{1}{|\mathcal{B}_{Tx}|} \sum_{k\in \mathcal{B}_{Tx}} |\hat{X}_i(k)|^2 \nonumber \\
  P_i[In] &=& \frac{1}{|\mathcal{B}_{In}|} \sum_{k\in \mathcal{B}_{In}} |\hat{X}_i(k)|^2.
\end{eqnarray}
We then define the \textit{transmit in port 0 to transmit in port 1} ratio (TTR) and the \textit{incident in port 1 to incident in port 0} ratio (IIR) as:
\begin{equation}
    \textrm{TTR}_{0 \rightarrow 1} =   \frac{P_0[Tx]}{P_1[Tx]}, \quad
    \textrm{IIR}_{1 \rightarrow 0} = \frac{P_1[In]}{P_0[In]}.
\end{equation}
Similar to signal to interference ratio (SIR), our TTR and IIR values measure a power ratio.  However, TTR and IIR do not require exact knowledge of the true transmitted and incident signals $X_0(k)$ and $X_1(k)$ for all $k$, which is unavailable, even during experiments. Our metrics, TTR and IIR, focus specifically on the isolation performance of \acronym{} rather than the quality of $\hat{X}_0(k)$ and $\hat{X}_1(k)$ individually.

\section{Implementation} \label{sec:impl}

In this section, we present the implementation of \acronym{} on a large-scale wireless testbed.  We first discuss the monitoring hardware, and specifications of their use in the experiments. Finally, the signal separation algorithm is given in Algorithm~\ref{alg:monitor}.

\subsection{Monitoring Hardware}

We use the following hardware in our experiments: (1) an SDR transmitter and incidental source each controlled via USB by an Intel NUC computer, (2) our custom bidirectional coupler connected between the TX port and antenna,  (3) the monitor node connected to the two output ports of the bidirectional coupler, and (4) a wide-band antenna \cite{nuc2021}. 
\begin{figure}[htb]
  \begin{subfigure}[b]{0.53\columnwidth}
    \includegraphics[width=\linewidth]{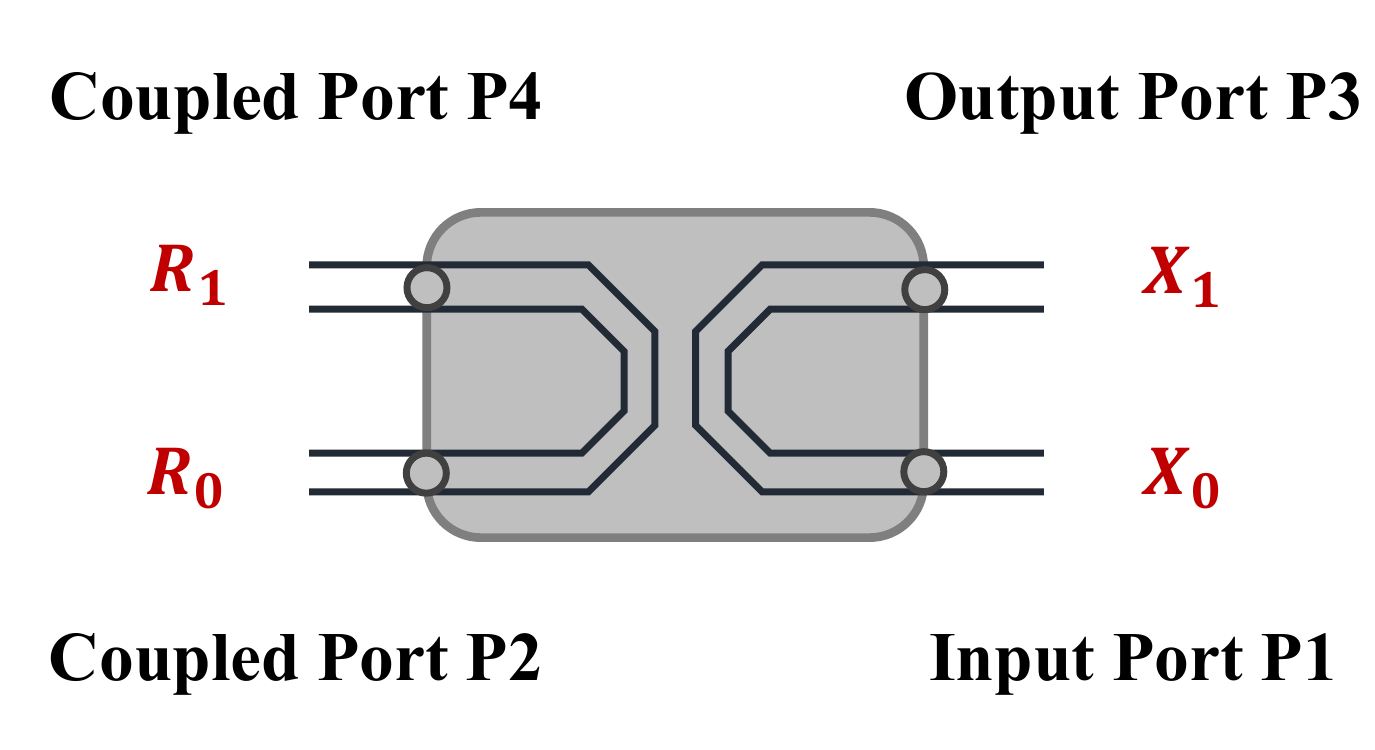}
    \caption{}
    \label{couplerdiagram}
  \end{subfigure}
  \hfill 
  \begin{subfigure}[b]{0.45\columnwidth}
    \includegraphics[width=\linewidth]{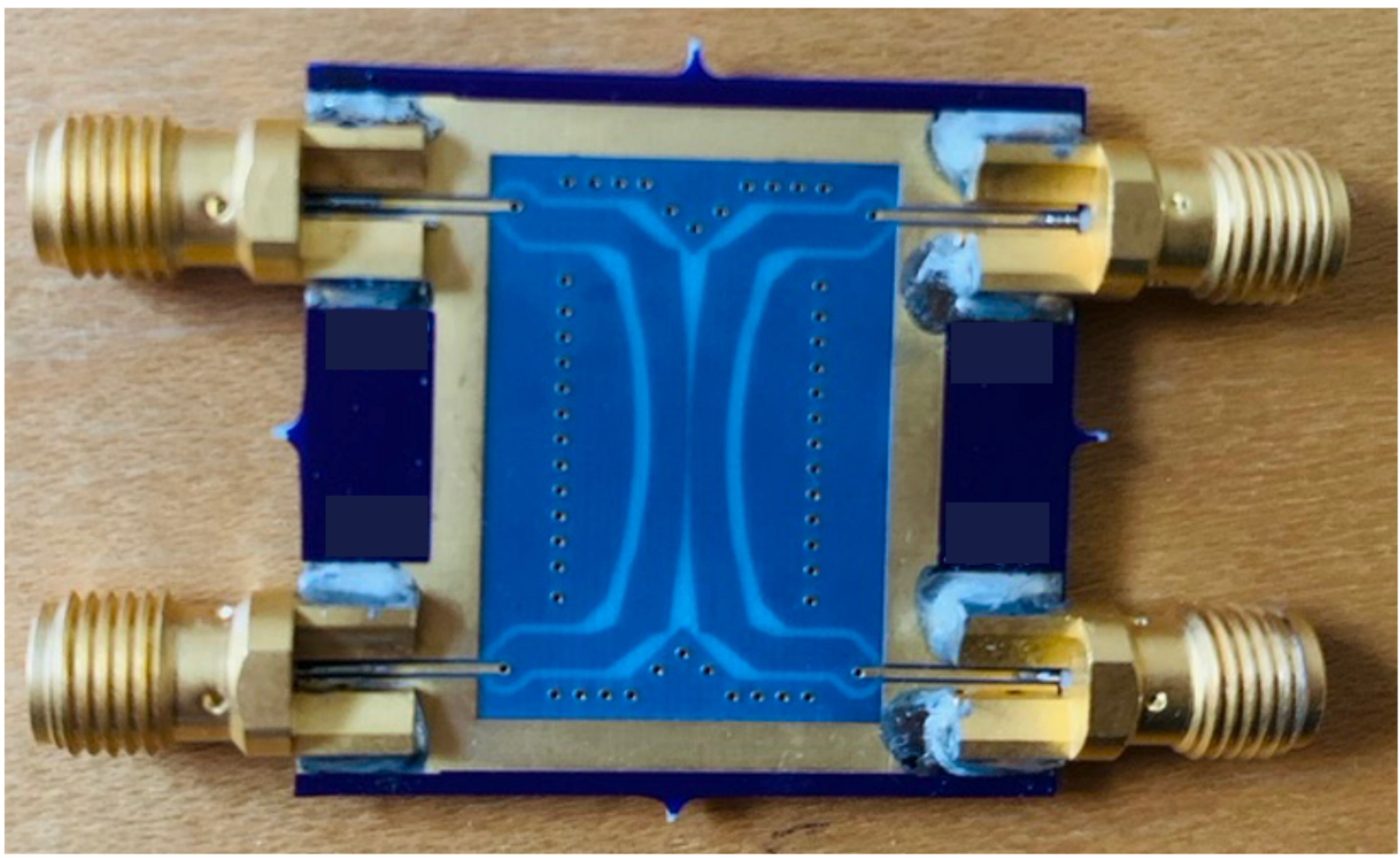}
    \caption{}
    \label{couplerPCB}
  \end{subfigure} \\
  \begin{center}
  \begin{subfigure}[b]{0.95\columnwidth}
  \includegraphics[width=\linewidth]{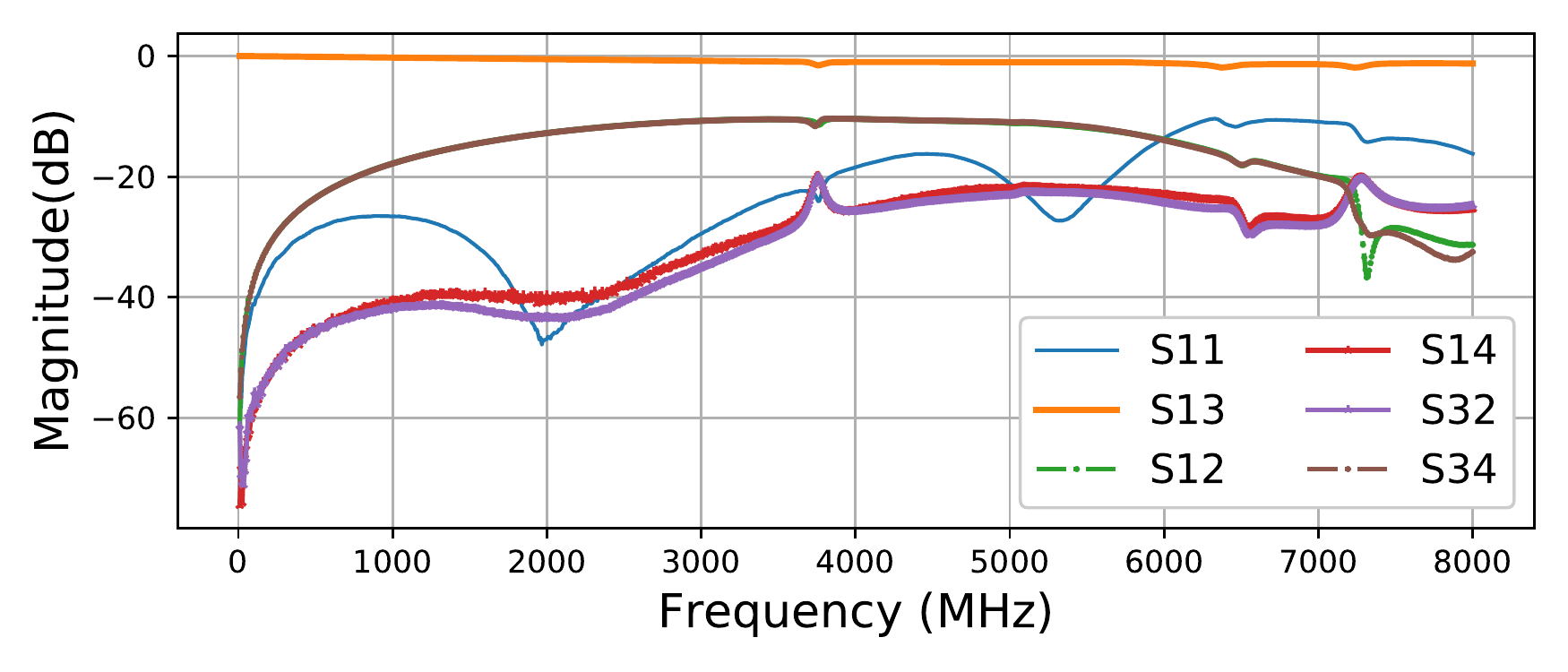}
    \caption{}
    \label{couplerS}
  \end{subfigure}
  \end{center}
\caption{
\label{coupler}
Bidirectional coupler (a) port diagram; (b) realization on PCB; (c) measured S-parameters.  While extremely wideband and low loss, the isolation between coupled ports can be as low as 10 dB.}
\end{figure}

We use a NUC, a small-form-factor PC with an Intel Core I7-8650 processor and 32~GB of DDR4 RAM, running Ubuntu 18.04 LTS. Our monitor and node SDR are both NI USRP B210s, which are able to transmit and receive in the spectrum range from 70--6000~MHz \cite{Ettus2015USRP}, with a sample rate up to 61~MSps. The antenna used is a TAOGLAS wide-band 4G LTE I-Bar, effective across a 698-6000~MHz band \cite{antenna2021}.

A critical component of the RF monitoring system, a bidirectional coupler, is designed and built as shown in Figure~\ref{couplerdiagram} and \ref{couplerPCB}. It has four ports: P1 and P3 are input and output ports representing direct transmission whereas P2 and P4 are coupled ports that can capture mixed signals at different scales. To show the directionality of the coupler when it is isolated, we measure its S-parameters across the 100--6000~MHz frequency range, as shown in Figure~\ref{couplerS}. $S_{11}$ shows low return loss, below $-10$ dB across the band. $S_{13}$ is close to 0 for the wide spectrum, indicating little power loss of the direct transmit signal from P1 to P3. In addition, $S_{12}$ and $S_{14}$ show that P2 and P4 receive a copy of the transmitted signal that is at least 10~dB and 20~dB down from the transmitted signal power, respectively. Likewise, $S_{32}$ and $S_{34}$ show that the samples collected in P4 and P2 are at least 20~dB and 10~dB down from the incident signal power.

\subsection{Monitoring Software} \label{E:ExperimentalSetup}

To monitor the entire 100--6000~MHz range, \acronym{} divides the spectrum into multiple smaller frequency channels. Issues with the NI B210 sometimes results in invalid samples when run at its maximum sampling rate, 
thus we use a sampling rate of 27.65~MSps, a factor of 0.9 of its maximum two-channel sampling rate. To cover the entire monitoring spectrum, 214 monitoring channels must be iterated through. For the results here, in each channel, $N=2\times 10^4$ complex-valued samples are collected by \acronym{} for source separation and mixing matrix estimation.

The \acronym{} procedure is detailed in Algorithm~\ref{RFMonitor}.
Note that PSD limits may be user-dependent, and we assume that they are known to the algorithm. Our implementation notifies the user and staff of the problem. Future implementations could be set to automatically shut off the transmitter.

\SetKwComment{Comment}{/* }{ */}
\begin{algorithm}[ht]
\caption{Algorithmic Operation of \acronym} \label{RFMonitor}
\SetAlgoLined
\KwResult{TX/incident signals, alert notification}
Initialize user PSD limits vs.\ frequency\;
Initialize the list of channel center frequencies $f_{\textrm{list}}$\;
\While{True}{
    \For{ $f$ in $f_{\textrm{list}}$ }{
        Sample $r_i(n)$ for $i=0,1$ \&  $n=0 \ldots N-1$ \;
        $R_i(k) \gets FFT\{r_i(n), i = 0,1\}$\Comment*[r]{Sec.\ref{model:FreqICA}}
        $\Tilde{\boldsymbol{X}}(k), \boldsymbol{\Tilde{A}} \gets \textrm{JADE} \{ \boldsymbol{R}(k) \}$\Comment*[r]{Sec.\ref{model:JADE}}
        $\hat{\boldsymbol{X}} \gets  \Delta(\boldsymbol{\Tilde{A}}) \Tilde{\boldsymbol{X}}(k)$\;
        $\hat{\boldsymbol{A}} \gets \left(\Delta(\boldsymbol{\Tilde{A}})\right)^{-1} \boldsymbol{\Tilde{A}}$ \Comment*[r]{Scaling}
        \eIf{$\textrm{corr}(|\hat{X_i}|, |\Tilde{R_m}|)>0.95 \, \forall i, m = 0, 1$}{
        $i \gets \operatorname*{arg\,max}_{i} (|\hat{X_i}|, \quad i = 0, 1$\\
        $m \gets \operatorname*{arg\,max}_{m} (|\Tilde{R_m}|), \quad m = 0, 1$\;
        }
        {$i, m \gets \operatorname*{arg\,max}_{i,m} \textrm{cor}(|\hat{X_i}|, |\Tilde{R_m}|)$\;}
        \If{$i \neq m$}{
            \(\hat{\boldsymbol{X}}(k) = \boldsymbol{J}_2 \cdot \hat{\boldsymbol{X}}(k)\)\;
            \(\hat{\boldsymbol{A}} = \boldsymbol{J}_2 \cdot \hat{\boldsymbol{A}}\)\Comment*[r]{Permutation}}
    }
   
    \textbf{Concatenate} $\hat{\boldsymbol{X}}(k)$ for all frequency channels $f$\;
    \textbf{Compute} PSD $= 10\log_{10} |\hat{X}_0(k)|^2$\;
    \If{PSD $>$ user PSD limits}{
        \textbf{Notify \& send} PSD graph to user \& staff\;
    }
}
\label{alg:monitor}
\end{algorithm}

\section{Results}\label{sec:result}

In this section, experimental results of \acronym{} are presented using the setup shown in Figure~\ref{Architecture}. 

\textbf{Baseline method.}
We use our preliminary work in~\cite{Terry2020spectrum} as a baseline to compare to \acronym{}. The baseline, \textit{System Matrix Calibration (SMC)}, works as follows.
SMC first calibrates the system matrix, at each frequency bin, by placing the system in a RF isolation area. In detail, a known signal is first transmitted by the experimental SDR and is received by the monitoring system at two ports. A spectrum analyzer captures the actual transmitted signal $X_0$. As the incident signal is zero, half of the mixing matrix can be estimated by comparing the measurements $\boldsymbol{R}$ and $X_0$. Similarly, by setting $X_0=0$ and transmitting a signal from an incident source (as measured by a spectrum analyzer), the other half of the mixing matrix is obtained. The calibrated linear model is then inverted and, during operation, is used to separate the transmitted and incident signals.



\subsection{Source Separation of \acronym{}}

Our first experiments are designed to answer the following critical question: \textbf{Can \acronym{} separate and identify signals of different modulations, center frequencies and bandwidths, and relative power levels?} 
We conduct multiple controlled experiments, in which we set the transmit signal, and create an environmental signal that impinges on the platform antenna, to answer this question.

\subsubsection{\textbf{Signal Type}}

\begin{table*}[t]
  \begin{tabular}{MMLLLLLL}
    \toprule
      \multicolumn{2}{c}{Types} &
      \multicolumn{2}{c}{Reference} &
      \multicolumn{2}{c}{{SMC}} &
      \multicolumn{2}{c}{{\acronym{}}}\\
    \hline
    transmitted signals  &  incident signals & TTR (dB) & IIR (dB) & TTR (dB) & IIR (dB) & TTR (dB) & IIR (dB) \\
    \midrule
    CW & OFDM & 13.30 & 7.21 & $\boldsymbol{30.92}$ & $\boldsymbol{3.54}$ & $\boldsymbol{30.95}$ & $\boldsymbol{19.60}$\\
   OFDM & CW & 12.56 & 6.91 & $\boldsymbol{22.61}$ & $\boldsymbol{7.39}$ & $\boldsymbol{29.60}$ & $\boldsymbol{18.14}$\\
    \midrule
    CW & BPSK & 13.38  & 6.02 & $\boldsymbol{30.24}$ & $\boldsymbol{4.48}$ & $\boldsymbol{30.26}$ & $\boldsymbol{16.33}$\\
   BPSK & CW & 12.55 & 6.40 & $\boldsymbol{20.06}$ & $\boldsymbol{8.32}$ & $\boldsymbol{28.06}$ & $\boldsymbol{18.65}$\\
   \midrule
    OFDM & BPSK & 13.12 & 6.33 & $\boldsymbol{22.64}$ & $\boldsymbol{5.60}$ & $\boldsymbol{29.68}$ & $\boldsymbol{16.36}$\\
   BPSK & OFDM & 13.17 & 7.76 & $\boldsymbol{20.48}$ & $\boldsymbol{3.85}$ & $\boldsymbol{28.48}$ & $\boldsymbol{19.48}$\\
    \bottomrule
  \end{tabular}
  \caption{TTR \& IIR for transmitted/incident signals of three types.  SMC increases signal isolation in CW scenarios only. \acronym{} provides more isolation and is effective across all signal types.}
  \label{tabel2}
\end{table*}

\begin{figure}[tbp]
    \centering
    \includegraphics[width=\columnwidth]{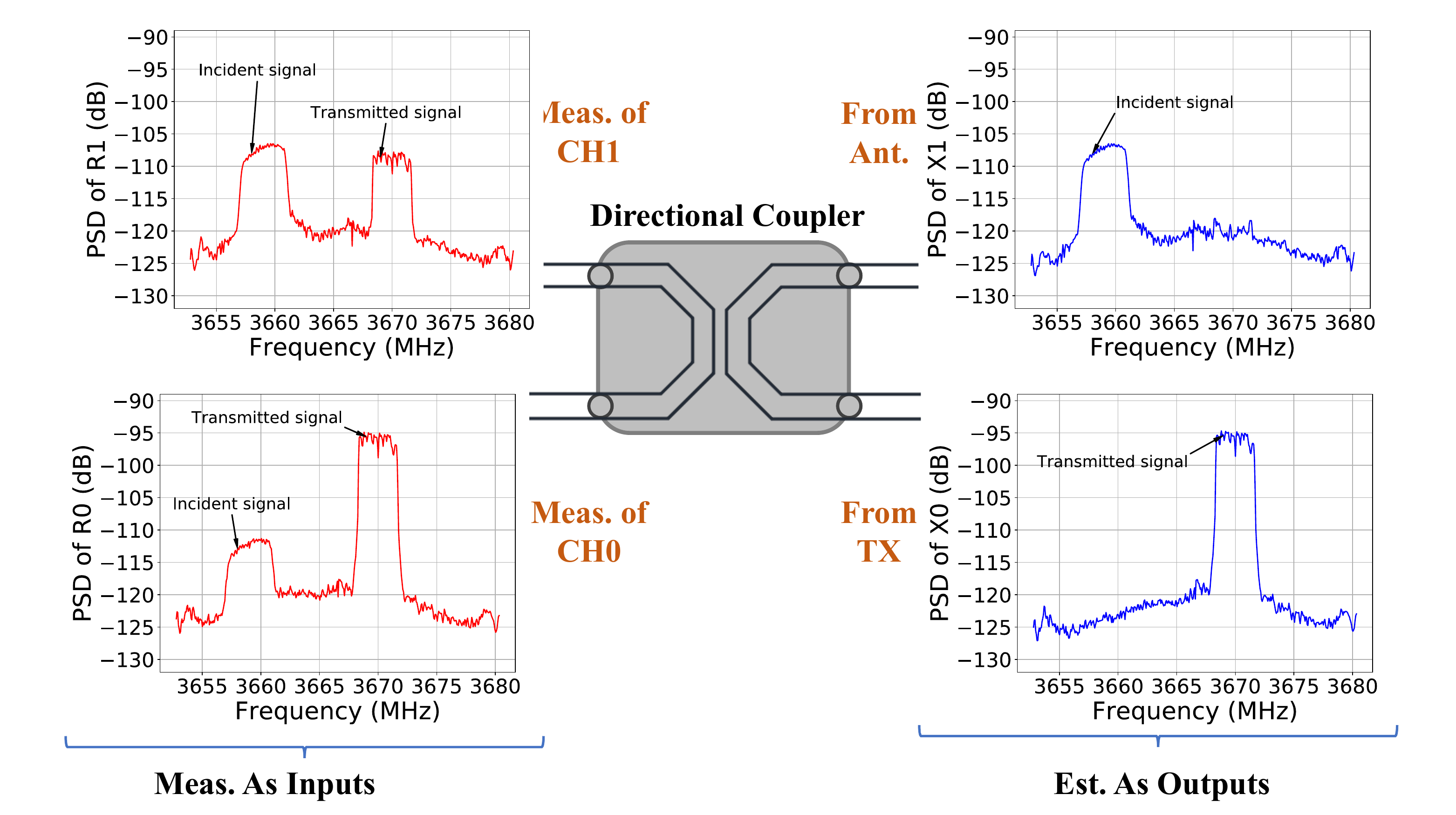}
    \caption{Different types of signals: \acronym{} almost completely separates an OFDM transmitted signal from BPSK incident signal. Improved TTR (13.12$\rightarrow$27.78~dB) and IIR (6.33$\rightarrow$17.36~dB).}
    \label{sep:types}
\end{figure}

First, we run an experiment in which an OFDM signal is transmitted at 3670~MHz and a BPSK signal is incident at 3659~MHz. Both have a bandwidth of 4~MHz. The results in Figure~\ref{sep:types} plot the PSD of $R_0$ and $R_1$ on the left, showing them at different power levels due to directionality of the coupler. The right plots of Figure~\ref{sep:types} show that only the correctly scaled transmitted signal remains in FDMonitor's output, $\hat{X}_0$. Likewise, the transmitted signal has been fully eliminated from the incident signal estimate $\hat{X}_1$.

We further compare \acronym{} to SMC with typical digital signal types, CW, BPSK, and OFDM, using TTR and IIR. The results in Table~\ref{tabel2} show that both methods increase isolation of the transmitted signal. However, we observe two disadvantages of SMC. First,
it cannot improve TTR as \acronym{} in CW transmission scenarios. Consider the first case with CW and OFDM: the reference TTR is 13.3~dB. SMC and \acronym{} enhance the isolation by similar amounts, 16.6 and 16.6~dB, respectively. However, large TTR differences in modulated transmitted signals expose the inability of SMC to remove the transmitted signal from $X_1$. When the transmitted signal is either OFDM or BPSK, the TTR increase via SMC is only 10~dB or 7~dB compared to 17 or 15~dB using \acronym{}. Secondly, SMC shows only a small IIR increase when the incident signal is CW, but inadvertently reduces the isolation for other signals. In the best case (row 4), SMC only increases IIR by 1.9~dB. In comparison, large IIR increases obtained using \acronym{} for all signal types indicate accurate and robust estimation of source signals.

\subsubsection{\textbf{Carrier and Center Frequency}}
\begin{figure}[btp]
    \centering
    \includegraphics[width=\columnwidth]{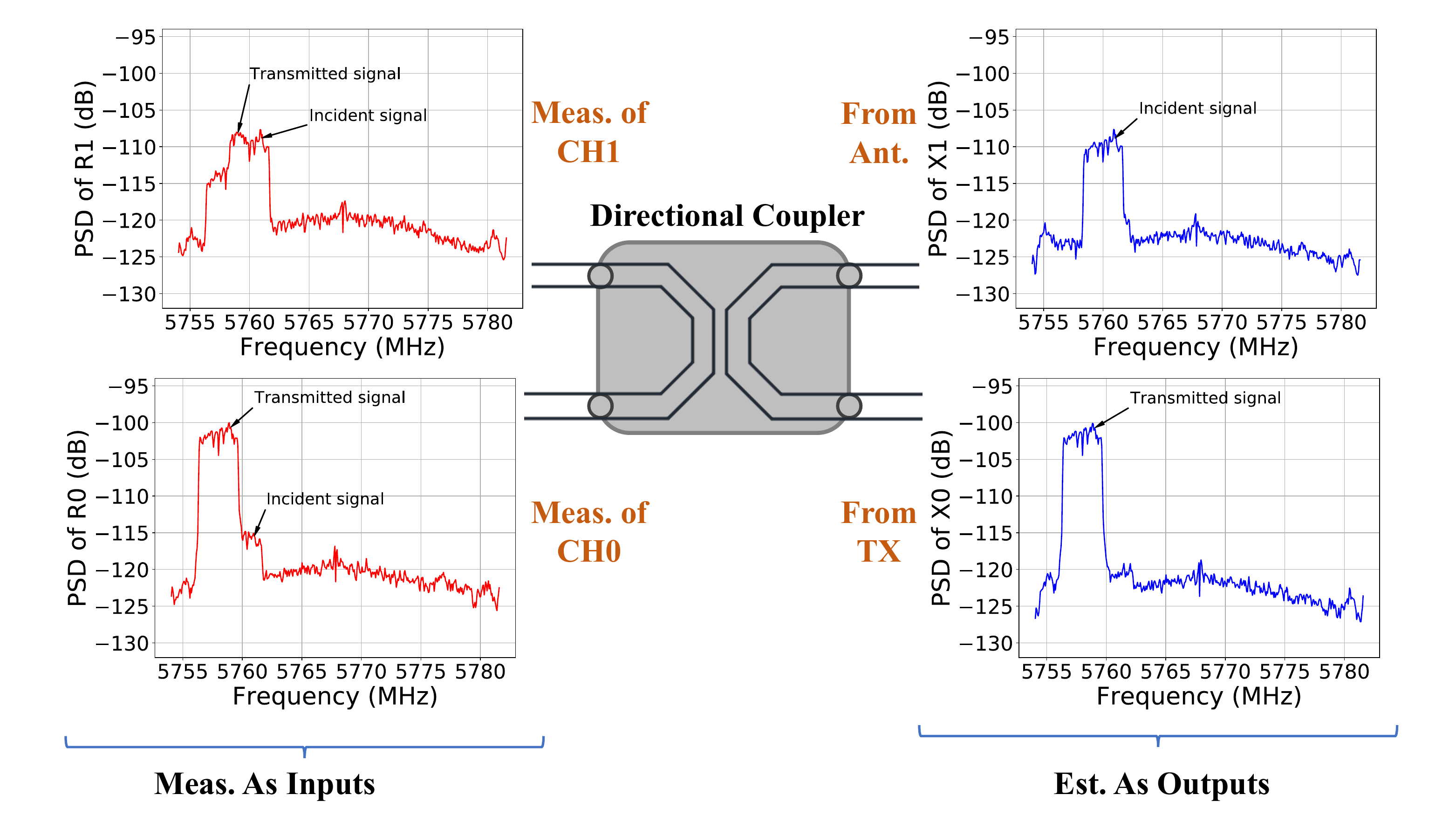}
    \caption{Frequency overlap: {\acronym } separates overlapping transmitted and incident OFDM signals. Only the incident signal remains in ${X_1}$, and only the transmitted in ${X_0}$.}
    \label{sep:carrier}
\end{figure}

Can \acronym{} separate signals that overlap in the frequency domain?  We conduct an experiment with two overlapping OFDM signals that overlap: an incident signal (5758--5762~MHz) and a transmitted signal (5756--5760~MHz) that overlap between 5758--5760~MHz. The  separation results in Figure~\ref{sep:carrier} show complete removal of the incident signal from ${X_0}$, and complete removal of the transmit signal from ${X_1}$. 

\begin{figure}[tbp]
    \centering
    \includegraphics[width=0.98\columnwidth]{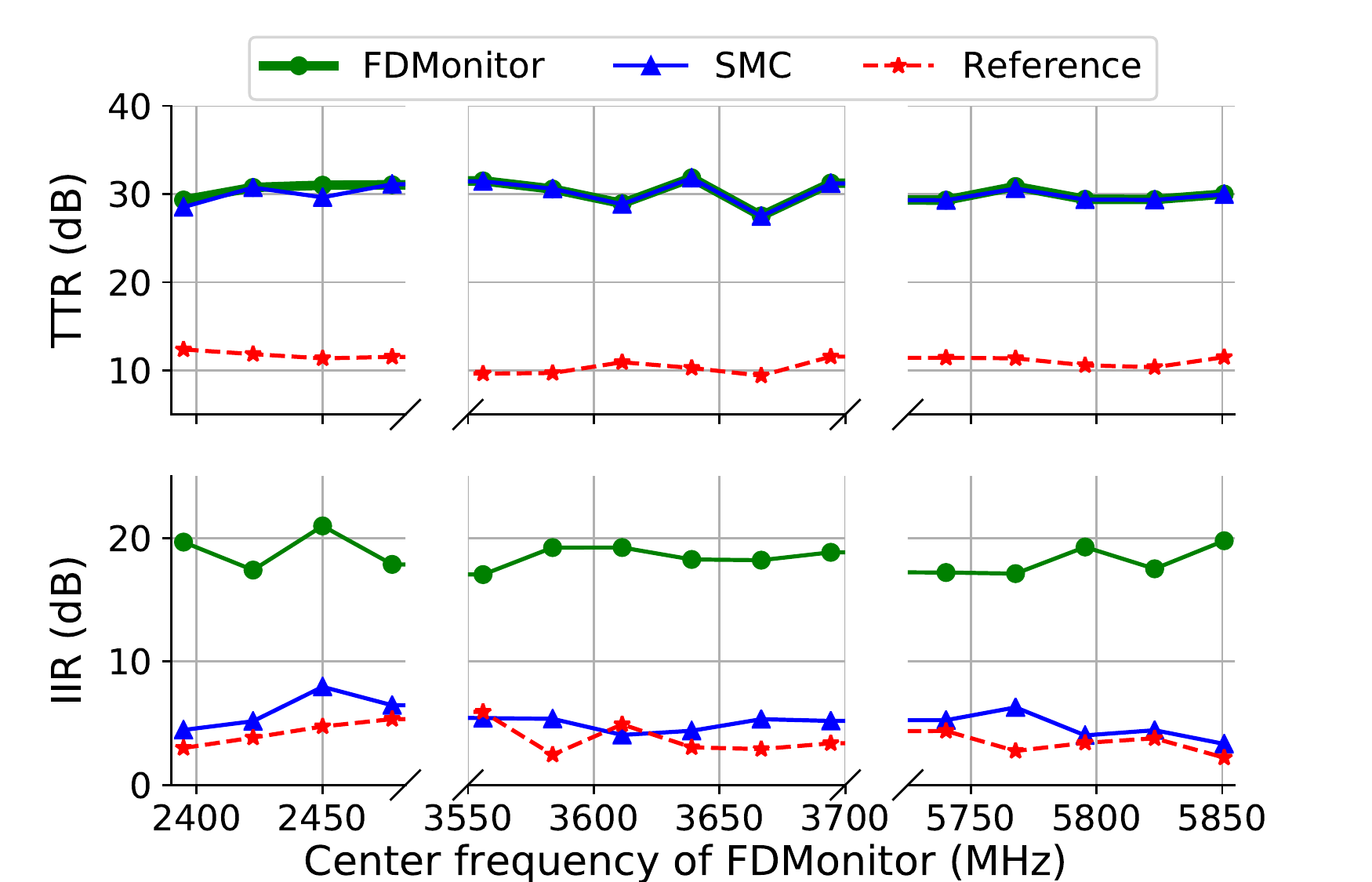}
    \caption{TTR and IIR  vs.\ frequency. Both methods can estimate the incident signal, but \acronym{} results in superior IIR.}
    \label{sep:centerfreq}
\end{figure}

To check the consistency of the separation performance across frequency at the receiver, experiments are conducted in the 2.4~GHz and 5.8~GHz ISM and 3.6~GHz CBRS bands, while transmitting non-overlapping CW signals. In Figure~\ref{sep:centerfreq}, we show that both methods increase the TTR across frequency, which means there is little impact of center frequency on either algorithm. Nevertheless, the large difference shown in IIR indicates poor incident signal separation of SMC and robust estimation via \acronym{}.

\subsubsection{\textbf{Signal Bandwidth}}

The transmitted and incident signal properties are unknown to \acronym{}, and might both occupy large bandwidths. We next run tests to explore how the performance of \acronym{} is affected by signal bandwidth.

\begin{figure}[htbp]
    \centering
    \includegraphics[width=\columnwidth]{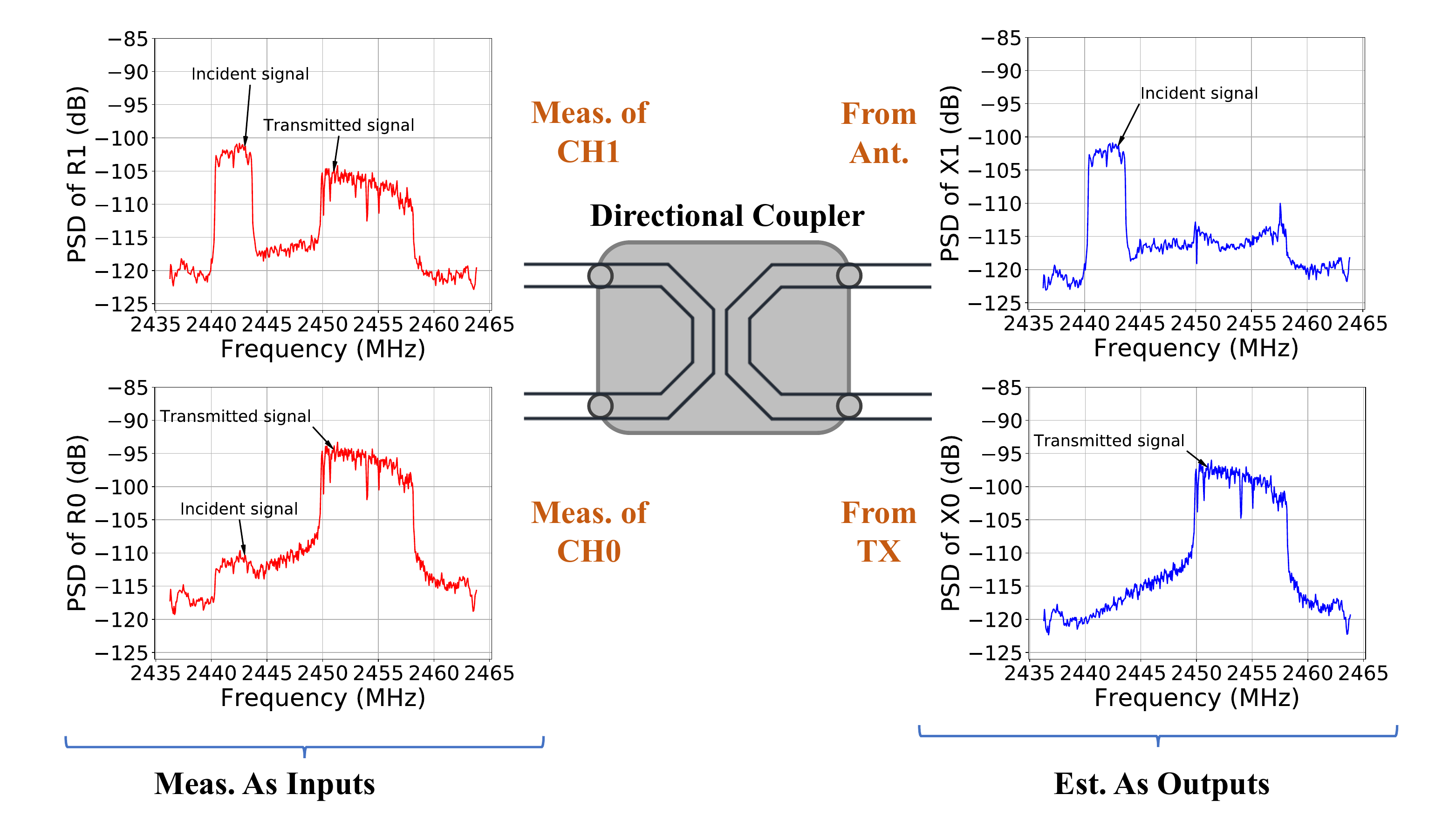}
    \caption{Different bandwidths: Signals of bandwidths 10~MHz \& 4~MHz are separated. 
    }
    \label{sep:bandwidth}
\end{figure}

In this experiment, we consider two OFDM signals: the transmitted signal is 10~MHz wide, centered at 2454~MHz, while the incident signal is at 2442~MHz with 4~MHz bandwidth. Figure~\ref{sep:bandwidth} shows the incident signal in $R_0$ is removed from $\hat{X}_0$. Equivalently, the transmitted signal has been mostly eliminated in $\hat{X}_1$ from $R_1$. Notably, 4~dB energy of the transmit signal on the edges remains in the $\hat{X}_0$. In addition, the spike at 2458~MHz has been confirmed (via separate investigation) as an environmental interference signal. This observation indicates that \acronym{} can perform source separation in the presence of multiple incident signals. 

\begin{figure}[htbp]
    \centering
    \includegraphics[width=0.95\columnwidth]{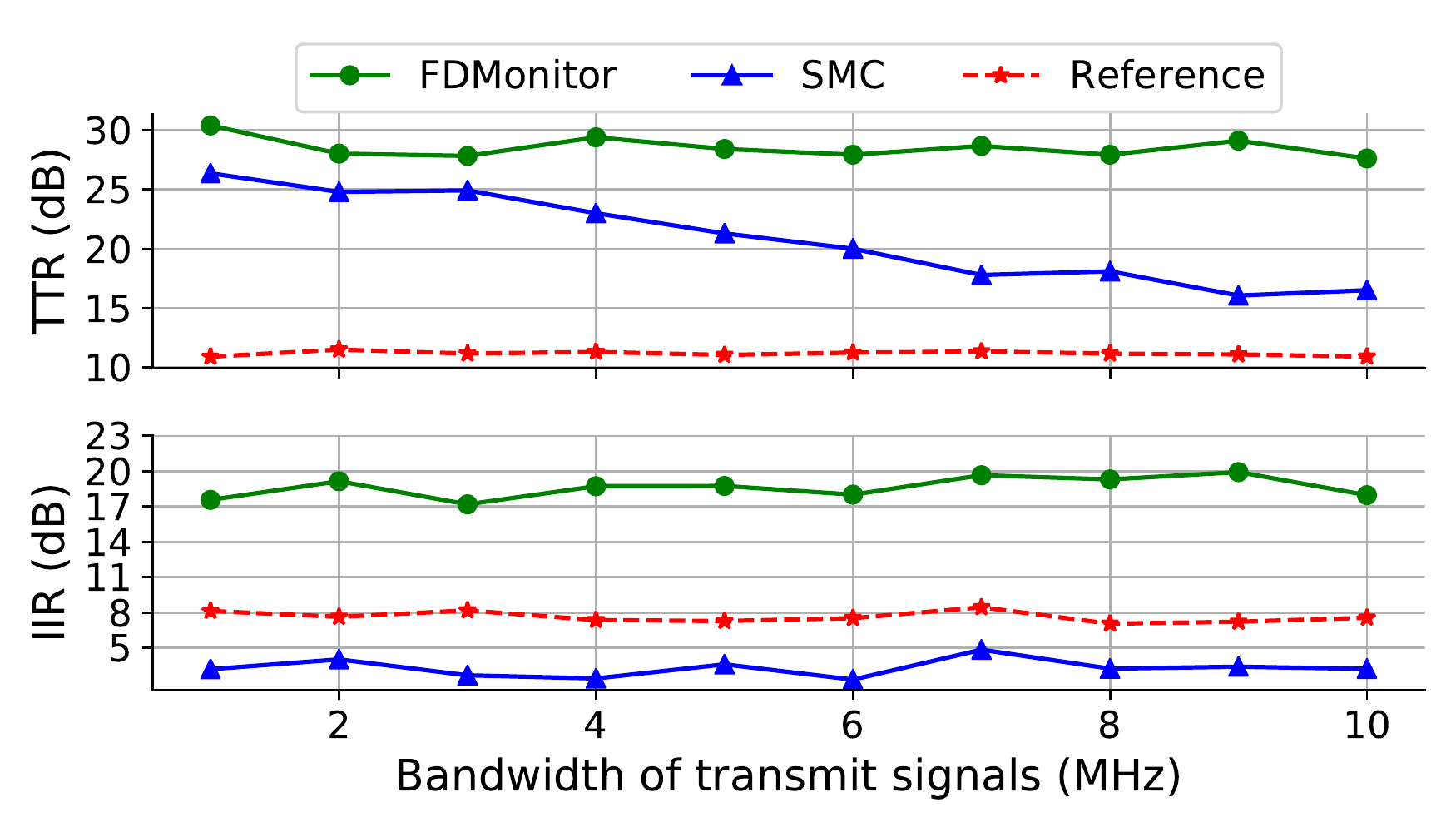}
    \caption{TTR and IIR variation vs.\ transmitted signal bandwidth. SMC degrades as bandwidth increases, whereas \acronym{} is consistent across bandwidth. 
    }
    \label{sep:bandwidthcompare}
\end{figure}

Figure~\ref{sep:bandwidthcompare} shows how TTR and IIR change as the transmit signal bandwidth varies from 1 to 10~MHz (10~MHz is the maximum bandwidth a user can reserve for one experiment on the platform under test). \acronym{} is stable across bandwidth, but SMC demonstrates decreasing TTR with higher bandwidth. Additionally, the IIR for SMC is stable but much lower than the IIR reference, whereas \acronym{} produces higher IIR. Reduced IIR means there is a higher level of incident signal in $\hat{X}_0$ than in $R_0$, a negative result.

\subsubsection{\textbf{Signal Power}}
Power difference as described in Section~\ref{S:ScalingAndPermutation}, can be used for solving permutation ambiguity. However, close power level may further confuse \acronym{}. Therefore, we present \acronym{}'s separation performance as a function of signal power.

\begin{figure}[tbhp]
    \centering
    \includegraphics[width=\columnwidth]{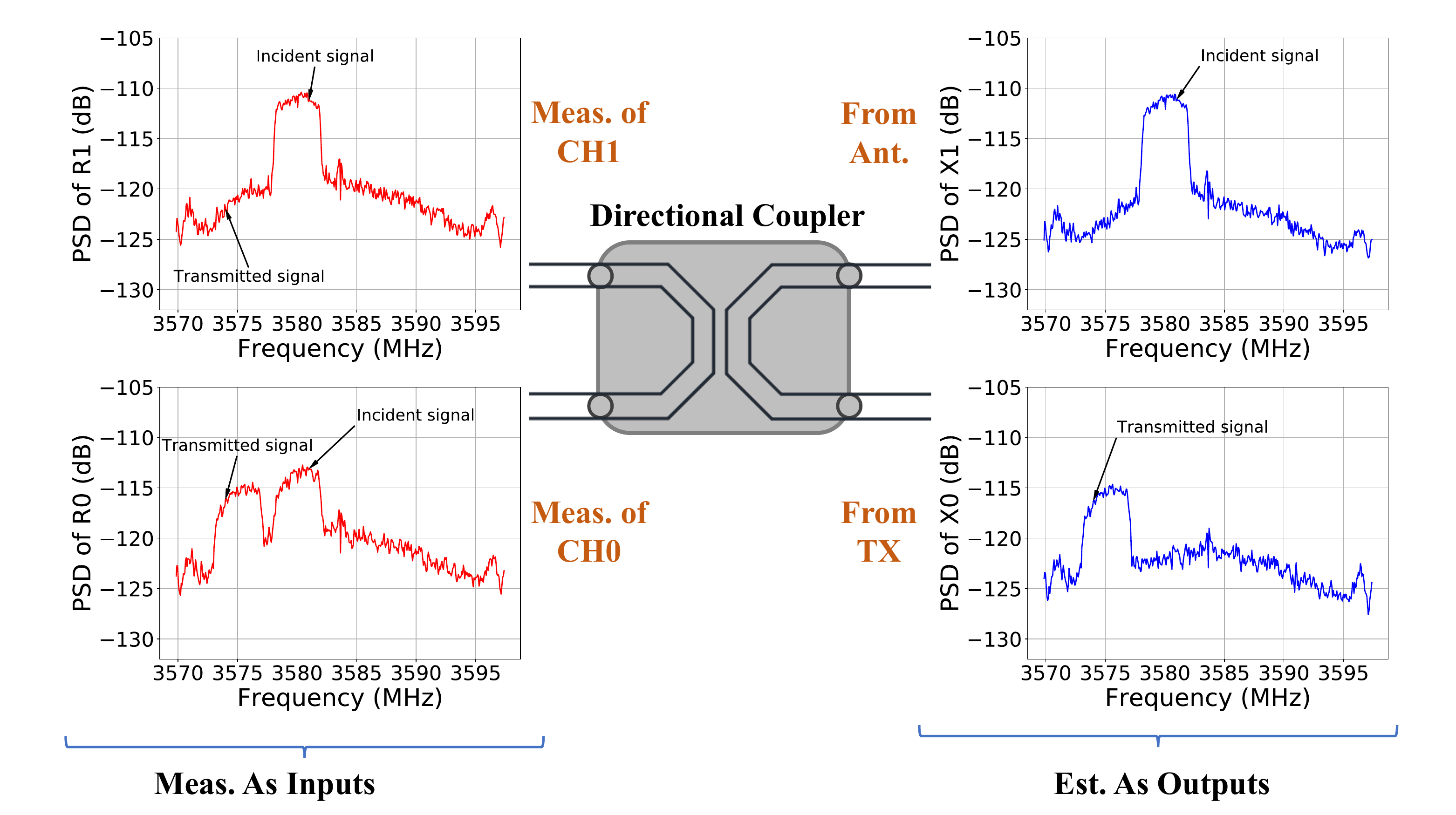}
    \caption{Similar Powers: \acronym{} separates BPSK signals despite similar power levels, $-115$ and $-113$ dB, increasing TTR ($5.9\rightarrow 9.6$ dB) and IIR ($2.1\rightarrow 10.6$ dB).}
    \label{sep:power}
\end{figure}

Figure~\ref{sep:power} shows the separation of two BPSK signals with close power levels, with the transmitted signal centered at 3575~MHz. As the transmitted signal gain is low, it is nearly invisible in the PSD of $R_1(k)$. Despite the low-power transmit signal, the results show  that \acronym{} accurately estimates and identifies each signal.

\begin{figure}[htbp]
    \centering
    \includegraphics[width=0.95\columnwidth]{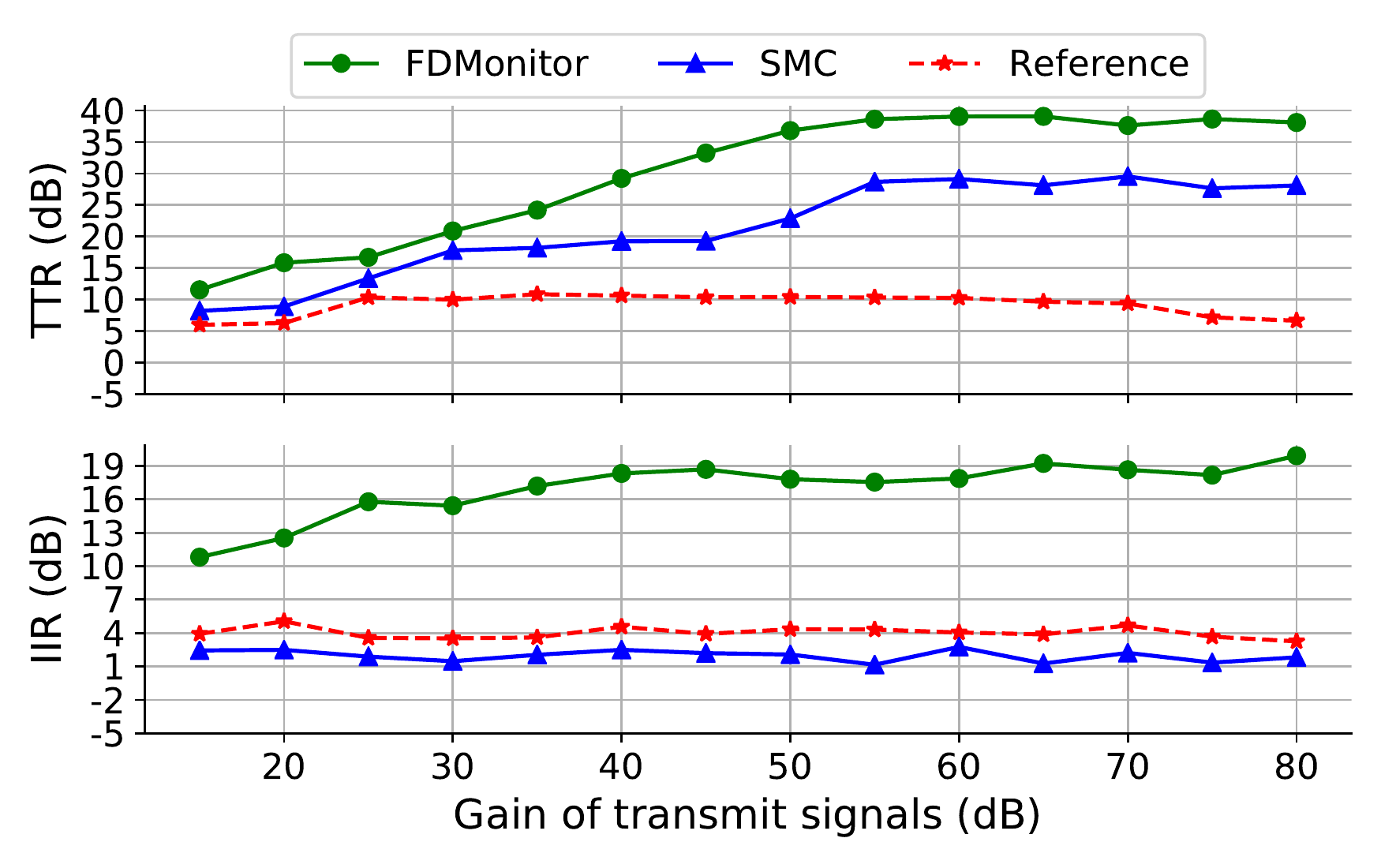} 
    \caption{TTR and IIR vs.\ transmitter gain:  Both improve with increasing gain in \acronym.}
    \label{sep:powercompare}
\end{figure}

Finally, we vary transmitter gain as shown in Figure~\ref{sep:powercompare}. The TTR and IIR for the reference method are approximately 10 and 5~dB. After running the two source separation algorithms, we make the following observations: (1) TTR for both methods increases while gain increases from 10 to 55~dB. (2) \acronym{} obtains higher TTR than SMC at each gain setting. (2) SMC provides IIR results around 2--3~dB lower than the reference, while \acronym{} increases IIR by 10-13~dB.

\subsection{Algorithm Efficiency}

\begin{figure}[tbp]
    \centering
    \includegraphics[width=0.7\columnwidth]{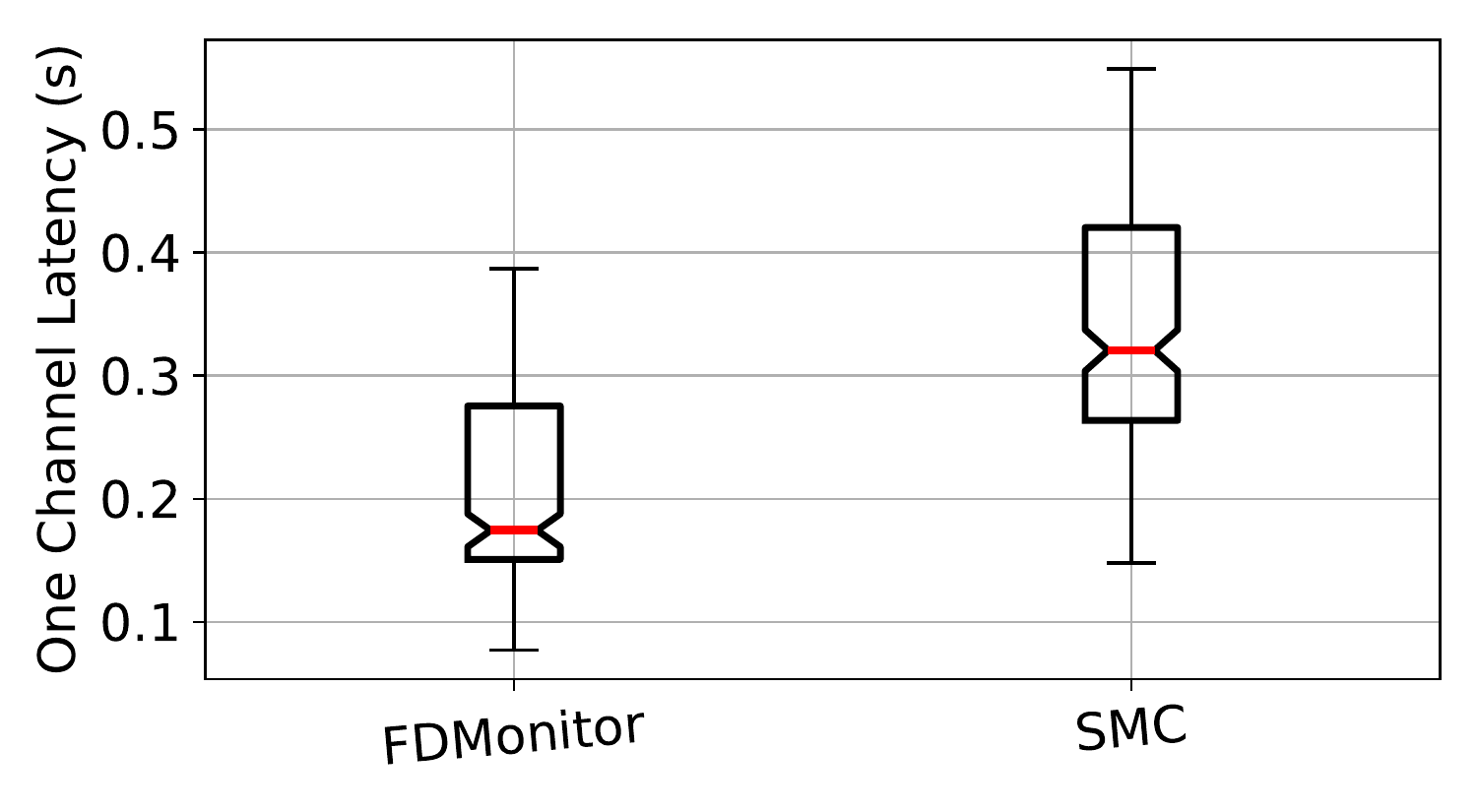}
    \caption{Latency of \acronym{} and SMC.}
    \label{latency}
\end{figure}

We compare the two systems' efficiency via latency. Latency in our work refers to elapsed time for frequency tuning, data collection and further analysis. Wideband monitoring of RF utilization involves two main components, frequency sweeping and source separation. The former tunes the center frequency of the receiver while the latter provides transmitted signal estimation in each 27.65~MHz sub-band. To assess the methods in terms of efficiency, latency is measured in each monitoring channel.  Figure~\ref{latency} shows that, for monitoring one channel, the median latency of \acronym{} is 0.17~s whereas SMC requires 0.32~s. Given the median latency, the total time spent by \acronym{} to sweep the 100-6000~MHz spectrum is 36.4~s, half the latency of SMC.

\subsection{Mixing Matrix Evaluation}\label{eval:matrix}
We further evaluate the system performance via the mixing matrix $\boldsymbol{\hat{A}} \in \mathbb{C}^{2\times 2}$, which describes the linear system model for source separation, which is estimated on the fly. We collect mixing matrix and precipitation data for 29 days while continuously transmitting CW signals from both sources. 

\begin{figure}[htbp]
    \centering
    \includegraphics[width=\columnwidth]{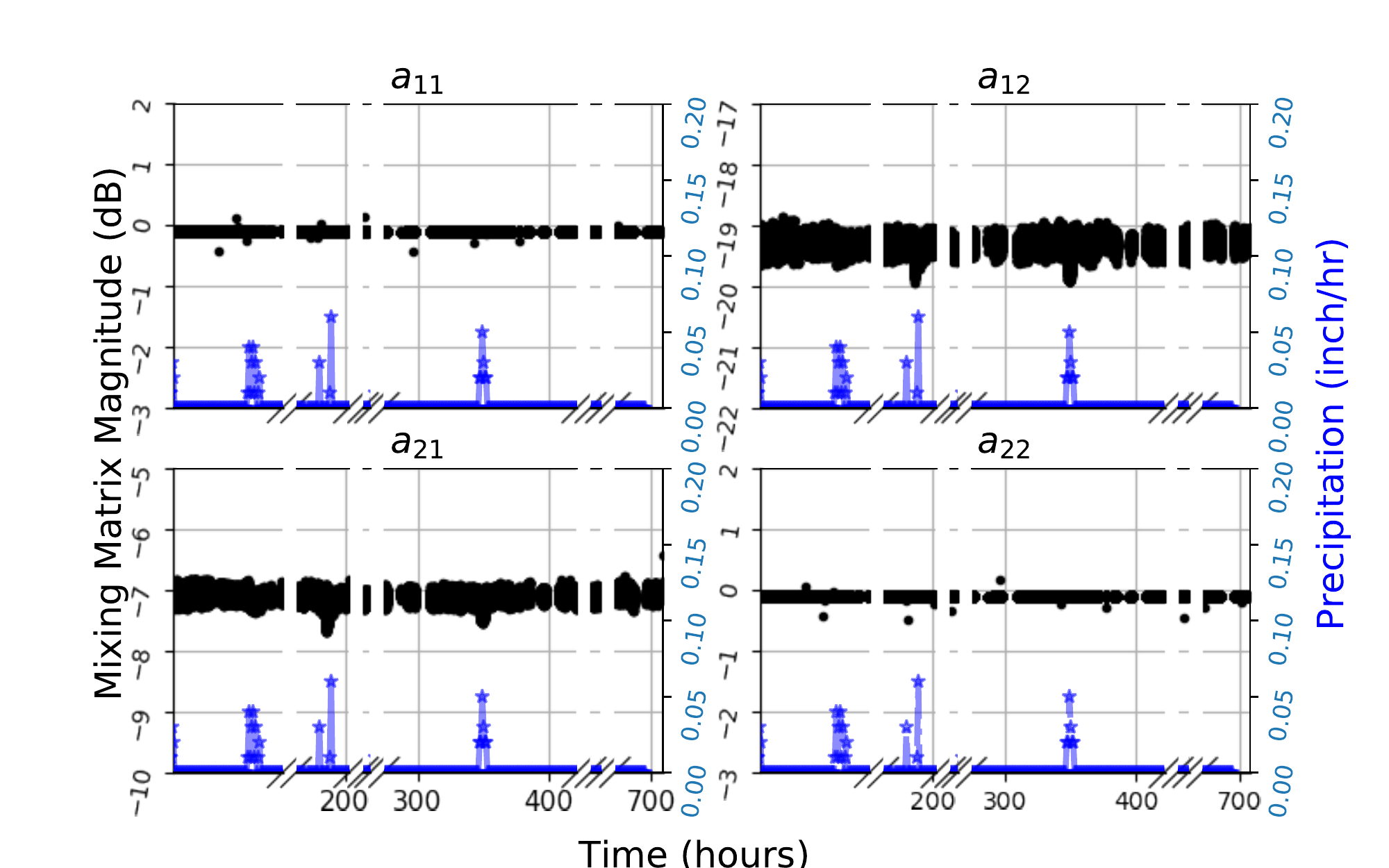}
    \caption{Mixing matrix magnitude over 105 hrs, with rain data. The magnitudes of matrix values are stable, but $a_{21}$ \& $a_{21}$ change while surfaces are wet due to rain.} 
    \label{fig15}
\end{figure}

Figure~\ref{fig15} shows the mixing matrix magnitude vs.\ precipitation and monitoring hours. First, we observe that the matrix magnitude is relatively stable across 29 days. Both $a_{11}$ and $a_{22}$ are centered at 0.01~dB with 0.002~dB standard deviation. Noisier $a_{21}$ and $a_{21}$ are around -19.29 and -7.10~dB respectively with 0.094 and 0.075~dB standard deviation. In addition, we notice that the matrix varies with rainfall. At around 72, 187 and 348 hours, the magnitude of $a_{21}$ and $a_{21}$ decreases when rain starts and later goes back to the same pre-rain level. This can be explained by the change in the radio propagation environment for wet vs.\ dry surfaces \cite{Hillyard2016Highly}. This is another reason why the baseline SMC method, which learns the system matrix only once during calibration, is not robust compared to \acronym{}.

\subsection{System-wide 7 Month Deployment}

\acronym{} has been continuously monitoring 19 SDR platforms on the testbed, all deployed at different geographical locations, for more than 7 months. We evaluate its system-wide performance by investigating each violation alarm it generates during the period, and considering the alarm accuracy.  A violation alarm notifies the user via email of detected signals being transmitted outside the declared spectrum. It can be a true detection of RF emission misbehavior, or a false alarm if the alert did not correspond to a user violating spectrum rules. We store measurements from each alert, and request information from the user about their setup in order to determine the ground truth about spectrum use.

The true violations so far, we classify into 5 types: 
\begin{enumerate}
    \item No declaration of any spectrum before using it,
    \item High gain setting induces harmonics, 
    \item Adjacent band spillover w/ wideband TX signals,
    \item 15~MHz intermodulation distortion, and 
    \item \acronym{} testing after installation.
\end{enumerate}


\begin{table*}[t]
  \begin{tabular}{c@{\hskip 0.3cm}M@{}M@{\hskip 0.5cm}M@{\hskip -0.2cm}M@{}L@{\hskip -0.2cm}M@{\hskip -0.2cm}L} 
    \toprule 
    Type & \multicolumn{2}{c}{False positives: 20 emails} & \multicolumn{5}{c}{True positives: 642 emails}\\
    \midrule
    Rate & \multicolumn{2}{c}{False discovery rate (FDR): 3.02\%} & \multicolumn{5}{c}{{Positive predictive value (PPV): 96.98\%}}\\
    \midrule
    Cause & Bug: Spectrum declaration lost & Permutation ambiguity & No spectrum declaration & High gain induced harmonics & Signal spillover & Intermod.\ distortions & System testing\\
    \midrule
    Rate & 65\% & 35\% & 67.29\% & 18.54\% & 0.16\% & 10.28\% & 3.74\%\\
    \bottomrule
  \end{tabular}
  \caption{\acronym{} alert accuracy during continuous monitoring of 19 shared SDR platforms for $>7$ months.}
  \label{monitor:violation}
\end{table*}

Table~\ref{monitor:violation} shows our analysis of the 662 total alerts.  In sum, we observe only 20 false alarms, among which 13 alerts occurred because the user-declared frequency was, due to a software error, not recorded to the \acronym{} user PSD limits database. The other 7 false discovery emails were triggered by incorrectly resolved permutation ambiguities. Even so, the $96.98\%$ positive predictive value (PPV) represents high accuracy and robustness of \acronym{} across a large variety of real users, their signals, and the varying weather seen by the platform.

\subsection{Adversarial Behavior}\label{threatmodel}

If \acronym{} sequentially monitors frequency channels to cover the entire 100-6000~MHz band, adversarial users can potentially transmit in violation while hopping between channels to avoid detection. To model adversarial behaviors, we use the following notation: 1) $\Delta T$ is the time duration to monitor one channel, 2) $N_C$ is the total number of channels (in our case, 214), and 3) $T$ denotes the cycle number, where each cycle corresponds to monitoring all $N_C$ channels.

\textbf{Attack model} We propose the following attack model: 1) an attacker can use any channel at any given time, 2) in each time slot $\Delta T$, an attacker chooses 1 of the $N_C$ channels to transmit, 3) an attacker does not know which channel is being monitored.

\begin{figure}[btp]
    \centering
    \includegraphics[width=\columnwidth]{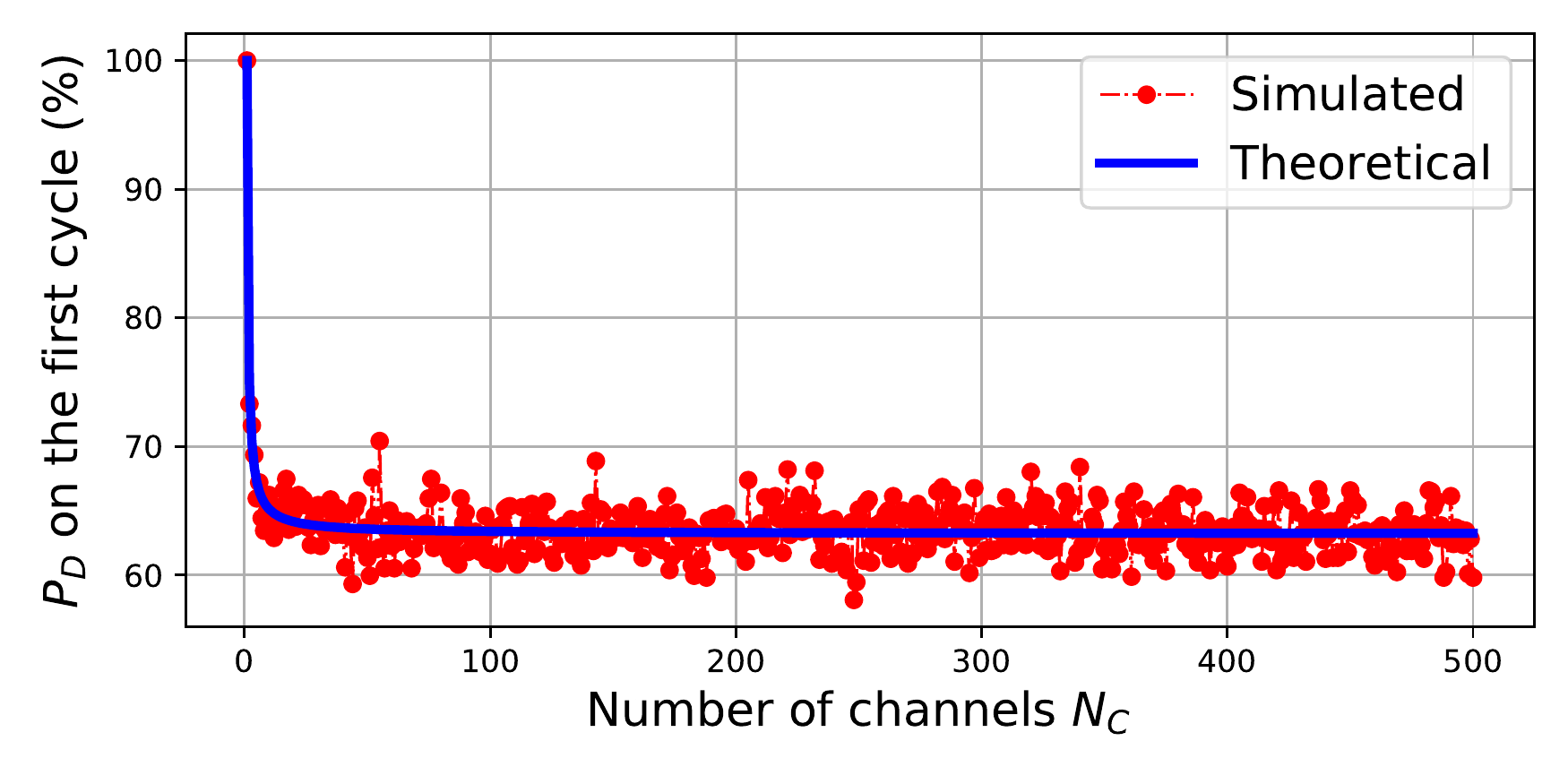}
    \caption{Prob. of detecting violation in the first cycle. In \acronym{}, $N_C=214$, and thus $P_D(1)=65.18\%$.} 
    \label{figure:countermeasureperformance}
\end{figure}

\textbf{Countermeasure} To address this attack model, \acronym{} can no longer use a predictable monitoring scheme. Instead, we propose a countermeasure that randomizes the order: 1) in each monitoring cycle, \acronym{} generates a random permuted channel sequence of length $N_C$ for spectrum monitoring, 3) all channels are measured by \acronym{} in each cycle. \textit{The probability of first detecting an attacker at cycle $T$ using the proposed countermeasure is}: 
\begin{equation}\label{eq:probability_of_detect}
    \begin{aligned}
    P_D(T) = \left(\frac{N_C-1}{N_C}\right)^{N_C(T-1)}\left(1- \left(\frac{N_C-1}{N_C}\right)^{N_C}\right).
    \end{aligned}
\end{equation}
$P_D(1)$ asymptotically converges to $1-\frac{1}{e}$ or 63\% as $N_C \rightarrow \infty$.  The average number of cycles for attacker detection is $1/P_D(1)$, which for high $N_C$ is 1.58 cycles. The proof is omitted due to space constraints. For validation, we show in Figure \ref{figure:countermeasureperformance} results of a simulation run $10^4$ times at each $N_C$.

We describe one type of adversarial behavior above. However, other adversarial behaviors are also possible. For instance, an adversary can use a particular channel for a duration longer than $\Delta T$. In this situation, the probability of detecting the adversary, $P_{D} (T)$ will actually be higher. In a different pathological scenario, an adversarial testbed user can have a colluder externally transmit the same disallowed signal that the user transmits, in a time-synchronized manner. If the external incoming signal is at sufficiently high power, \acronym~can miss detecting the user signal. We will investigate such pathological adversarial cases in our future work.

\section{Discussion}\label{discussion}

We describe some limitations of \acronym{} and discuss the implications for future research. 

\textbf{Non-Gaussian constraint} One constraint of \acronym{} is Assumption~\ref{as:sources} that at most one of the sources is Gaussian.  One possible solution is that, if both signals are found to be Gaussian, \acronym{} could use a recently estimated system matrix $\boldsymbol{\hat{A}}$ for separation instead of estimating it on the fly.

\textbf{Generalization to MIMO platforms} Generally, one \acronym{} is needed for each transmit antenna, since it  currently relies on a bidirectional coupler to separate the transmit and incident signal to that antenna. This could be a scaling problem for MIMO platforms. However, future work could use a $(N+1)$-directional coupler for monitoring $N$-antenna MIMO platforms assuming that the incident signal appears in different linear combinations on all antennas. By doing so, $N+1$ rather than $2N$ measurements would be needed for MIMO ICA separation.

\section{Conclusion}\label{sec:conc}
This paper proposes, implements, and reports on \acronym{}, a robust and continuous full-duplex monitoring system for effective supervision of a SDR platform's transmissions and environmental use of spectrum. \acronym{} uses a novel frequency-domain source separation algorithm to distinguish signals of the SDR platform from those incident on the antenna. Critically, our approach does not require extensive calibration, which would be very challenging to implement at the rate at which calibration becomes obsolete. 
Its performance is extensively validated with four different types of RF signal experiments, across  communication signal types, carrier frequency, bandwidth and transmit power. We evaluated more than 7 months of live system performance, which generates a low $3.02\%$ FDR, with 20 false alerts.

\section{Acknowledgment}
This research is supported in part by the US National Science Foundation Grants 1564287 and CNS-1827940, and the PAWR Industry Consortium.

\bibliographystyle{plain}
\bibliography{FullduplexMonitor}

\end{document}